\documentclass[nobibnotes, nofootinbib, aps, pra, twocolumn]{revtex4-2}
\usepackage{graphicx}
\usepackage{amssymb}
\usepackage{amsmath}
\usepackage{enumitem}
\usepackage{subcaption}
\usepackage{braket}
\usepackage{multirow}
\usepackage{tikz}
\usetikzlibrary{quotes, angles}
\usepackage{tikz-3dplot}

\begin{document}

\title{Acceleration Noise Induced Decoherence in Stern-Gerlach Interferometers for Gravity Experiments}

\author{Meng-Zhi Wu}
\email{mengzhi.wu@rug.nl}
\affiliation{Van Swinderen Institute for Particle Physics and Gravity, University of Groningen, 9747 AG Groningen, the Netherlands }

\begin{abstract}
    Stern-Gerlach interferometer (SGI) is a kind of matter-wave interferometer driven by magnetic field and has been proposed for various gravity experiments. Stochastic noises can lead to decoherence problems of SGI via various mechanisms. In this paper, I will theoretically study several mechanisms including dephasing, contrast loss and position localisation decoherence.
    I will firstly present a rigorous proof that the dephasing behaves as a linear response to the noise, with a transfer function given by the Fourier transform of the unperturbed classical trajectories.
    Then I will demonstrate that stochastic acceleration noise only induces dephasing to the witness operator constructed in spin space, while it does not lead to contrast loss or position localisation decoherence due to common mode cancellation. In contrast, higher-order noise can induce both contrast loss and position localisation decoherence, contributing a decay factor proportional to the noise power spectrum density at the intrinsic frequency, which can be physically interpreted as the resonance between the noise and test mass. Based on the result, I apply the framework to analyse two typical noise sources, magnetic field noise and quadratic noise. 
    
\end{abstract}

\maketitle

\section{Introduction}\label{section 1}

Stern-Gerlach interferometer (SGI) is a type of matter-wave interferometer driven by an external magnetic field through the spin of the system~\cite{Keil2021}. In recent years, SGIs based on Nitrogen-Vacancy (NV) centres~\cite{DOHERTY20131} has been proposed as gravity sensors~\cite{Zhao2022InertialMW} like gravimeters~\cite{PhysRevLett.111.180403, Chen:18} and gyroscopes~\cite{PhysRevA.86.052116, PhysRevA.86.062104}. They have also been proposed for exploring the nature of gravity~\cite{PhysRevLett.119.240401}.  
Compared to the matter-wave interferometers driven by lasers, the test mass of an SGI is usually much heavier. Consequently, the amplitude of the signal of gravitational interactions can be significantly larger and easier to detect. 

Robustness is an essential issue for all types of matter-wave interferometers, including the SGI. In real experiments, 
test masses are affected by acceleration noises arising from various sources, such as mechanical vibrations~\cite{hfxv-c598}, inertial forces~\cite{PhysRevResearch.3.023178} and Coulomb interations~\cite{PhysRevA.110.022412}. The test masses also couple to higher order noises including gravity gradient (tidal force) noise~\cite{PhysRevResearch.3.023178}, rotation noises~\cite{PhysRevD.111.064004} and magnetic field noise~\cite{dbrs-wn92, 9n6y-cc7r}. 
These noises can disturb the trajectories and velocities of the two arms of the interferometer at the classical level and can cause decoherence of the quantum state. 
Since noises are usually characterised by their power spectrum densities (PSD), a fundamental question arises: how to formulate the decoherence factors caused by noises based on their PSDs?

Numerous previous studies have investigated the decoherence of SGIs, classifying it into several typical mechanisms including the \emph{dephasing}, the \emph{position localisation decoherence}~\cite{Joos1985, PhysRevA.84.052121, Romero-Isart_2017, Pino_2018, PhysRevLett.132.023601, RieraCampeny2024wigneranalysisof} and the loss of contrast (known as the \emph{Humpty-Dumpty problem})~\cite{Englert1988, Schwinger1988, PhysRevA.40.1775}. 

The dephasing effect arises from the ensemble average of random phase fluctuations induced by stochastic noise. It can be classified into the phase fluctuation of the spin phase~\cite{PhysRevA.97.012104, 10.1063/1.5089550, doi:10.1080/00018730802218067} and the spatial path integral phase~\cite{PhysRevResearch.3.023178, PhysRevD.107.104053, PhysRevA.110.022412, PhysRevD.111.064004}. In spin space, dephasing is typically characterised by the $T_2$ time in the Bloch-Redfield equation, which is a linear response to the PSD of the noise~\cite{10.1063/1.5089550, doi:10.1080/00018730802218067}. As for the spatial dephasing, some researchers have observed that it is also a linear response to the noise, characterised by a transfer function equal to the Fourier transform of the classical unperturbed trajectories~\cite{PhysRevResearch.3.023178, PhysRevD.107.104053, PhysRevA.110.022412, PhysRevD.111.064004}. Nonetheless, several subtle gaps remain to be proved: (i) why does the dephasing depend solely on the classical trajectories of the centre of mass; (ii) why does the fluctuation of the trajectories provide no contribution to the dephasing?

Position-localisation decoherence, which pertains to the decoherence of the spatial wavefunctions, was originally studied to elucidate the collapse model~\cite{Joos1985} of quantum states caused by the interactions with the ambient environment. This type of decoherence is typically attributed to scattering processes involving particles such as air molecules or photons~\cite{Joos1985, schlosshauer2007quantum, PhysRevA.84.052121}. Furthermore, the ensemble average of stochastic noises can also induce a dissipator $\mathcal{D}[\cdot]=-\Lambda[\hat{x}, [\hat{x}, \cdot]]$, leading to a decay factor $\mathrm{e}^{-\Lambda(x_1-x_2)^2t}$ applied to the spatial density matrix~\cite{VANKAMPEN1974239, VANKAMPEN1976171, Pino_2018, RieraCampeny2024wigneranalysisof}. In a scenario involving SGI, the witness operator is usually constructed within the spin space, prompting a question how the position localisation decoherence affects the spin witness. 

The Humpty-Dumpty problem is a peculiar issue in SGI~\cite{doi:10.1126/sciadv.abg2879, Keil2021}, while the former two effects occur in all types of interferometers. In this type of interferometers, the witness is usually constructed in the spin space, necessitating to trace out of spatial degrees of freedom. Consequently, the contrast loss can lead to a decay factor of the off-diagonal terms in the spin density matrix, known as the Humpty-Dumpty problem. The most fundamental source of the contrast loss arises from the non-closure of the two arms of the interferometer at the final time~\cite{Schwinger1988, PhysRevA.104.053310, doi:10.1126/sciadv.abg2879}. Other plausible sources have been theoretically studied in recent researches, such as the rotation effect~\cite{PhysRevLett.130.113602}, the coupling with phonons~\cite{10.1116/5.0080503} and the deformation of the wave-packet~\cite{Zhou:2024voj}. Notably, classical noises can also cause contrast loss problem, warranting further investigation to formulate the associated contrast loss. 

It is evident that all these mechanisms are intrinsically linked to the PSD of the noise, but previous papers have studied them separately. It underscores the need for systematic research that integrates all these mechanisms via the full dynamics of the test mass, despite mathematically challenging posed by the time dependence and randomness of noises. Consequently, most previous researches involving the full dynamics have focused on the white noise~\cite{Chen:18, Band_2022, Rong2015}, which is identified by a constant PSD and lacks time correlation. Some authors have noticed the limitations of analysing white noise, prompting them to consider the time dependence of the noise and to solve the corresponding Schrödinger equation using perturbation methods~\cite{PhysRevLett.125.023602}. However, the connection between the decoherence factor and the PSD remains unestablished.

In light of the aforementioned issues, this paper aims to formulate the decoherence mechanisms outlined above based on the PSD of the external noise. The system will be modelled as a harmonic oscillator under a driven force with fluctuations, and its time evolution can be analytically solved using an elegant method known as the \emph{Lewis-Resenfeld invariant}~\cite{lewis1969exact, 10.1063/1.525329, MIZRAHI1989465, Guerrero_2015} and \emph{quantum Arnold transformation}~\cite{Aldaya_2011}. 

Building upon the exact dynamics, I shall present a systematic analysis on the aforementioned decoherence mechanisms in the presence of an acceleration noise. In particular, the following topics will be addressed:
\begin{itemize}[leftmargin=8pt, noitemsep]
\item I begin with a rigorous proof demonstrating that the dephasing is as a linear response to the noise, followed by an investigation on mathematical properties of the associated transfer function. 
\item Next, I examine the impact of position localisation decoherence and the Humpty-Dumpty problem induced by acceleration noises. As will be demonstrated, this category of noise does not lead to loss of spin witness, which can be interpreted as a form of common mode cancellation~\cite{PhysRevLett.81.971, PhysRevA.89.023607, WANG201634}. 
\item Third, I offer a preliminary discussion of the leading-order effects of higher-order noise processes, analysing the loss of witness arising from these mechanisms.
\item Finally, I consider two representative noise sources, spin-dependent noise and trap frequency noise (i.e. quadratic noise), as applications of the general framework developed for noise-induced decoherence.
\end{itemize}

This paper is organised as follows. Section 2 reviews the dynamics of an ideal Stern-Gerlach interferometer (SGI) in the absence of noise, with subsections 2.1 and 2.2 addressing the spatial motion and spin-space witness construction, respectively. Mathematical details are provided in Appendices A and B. Section 3 focuses on the dephasing effect, while Section 4 explores the position localisation decoherence and the Humpty-Dumpty problem induced by acceleration noise. Section 4 is divided into several subsections, each addressing distinct aspects: the fluctuation of the classical trajectories, the fidelity loss of a single arm, position localisation decoherence, and the loss of spin witness. An important integral is computed using the residue theorem in Appendix C, while the master equation for single-arm localisation decoherence is derived in Appendix D. Section 5 presents a preliminary discussion on the witness loss induced by higher-order noises, while Section 6 applies the results to analyse two typical noises, spin-dependent noise and quadratic noise, each treated in a separate subsection. The analytic solution for the quadratic noise is summarised in Appendix E. Finally, Section 7 concludes with remarks on the role of randomness (7.1) and a summary of the paper (7.2).

\section{Ideal Stern-Gerlach Interferometer without Noises}\label{section 2}

\begin{figure*}
    \centering
    \begin{tikzpicture} 
        \tikzstyle{ball} = [circle, ball color=cyan, shading=ball, minimum size=10mm, inner sep=0pt]
    
        \draw[->, gray, opacity=0.8] (-1.5,-3) -- (9,-3) node[right, black]{$t$}; 
        \draw[->, gray, opacity=0.8] (-1.5,-3) -- (-1.5,3) node[right, black]{$x$};
        \draw[->, line width=0.75mm, blue] (-2, -2, 0) -- (-2, 2, 0) node[left] {$B(t)=\eta(t)x$};

        \node[ball, opacity=0.8] at (0,0) {};
        \node[ball, opacity=0.4] at (4,2) {};
        \node[ball, opacity=0.4] at (4,-2) {};
        \node[ball, opacity=0.8] at (8,0) {};
        \draw[domain=0:8,smooth, line width=0.5mm, teal] plot (\x, {2*sin(3.14*\x/8 r)});
        \draw[domain=0:8,smooth, line width=0.5mm, teal] plot (\x, {-2*sin(3.14*\x/8 r)});
        \draw[->, line width=0.75mm, orange] (0,0) -- (0,-0.8) node[below] {$mg$};
        \draw[->, line width=1mm, red] (4,1.65) -- (4,2.35);
        \draw[->, line width=1mm, red] (4,-1.65) -- (4,-2.35);
        \node at (0.1, -1.5) {$\ket{S}=(\ket{\uparrow}+\ket{\downarrow})/\sqrt{2}$};
        \node at (8.5, -1.5) {$\ket{S}=(\ket{\uparrow}+\mathrm{e}^{i\phi_{\rm diff}}\ket{\downarrow})/\sqrt{2}$};
    \end{tikzpicture}
    \caption{\small The schematic diagram of an ideal Stern-Gerlach interferometer for gravity experiments. Under a non-uniform magnetic field $\mathbf{B}=\eta\mathbf{x}$, the test mass splits to two trajectories with respect to different spin states. The magnetic field gradient $\eta\equiv\partial B/\partial x$ can be set as a time-independent constant~\cite{PhysRevLett.111.180403}, a piecewise function of time~\cite{PhysRevLett.125.023602}, several pulses~\cite{doi:10.1126/sciadv.abg2879}, or more complicated functions~\cite{PhysRevResearch.4.023087, Zhou:2024voj}. The differential phase $\phi_{\rm diff}$ at the final time encodes the information of gravity, so properties of gravity can be detected by measuring this phase.}
    \label{schematic}
\end{figure*}

A general schematic diagram of a Stern-Gerlach interferometer is shown as Fig.\,\ref{schematic}. The test mass is trapped in a potential well which can be assumed as a harmonic oscillator for simplicity. Besides, the spin of the test mass has to be split during the experiment, which can be achieved by some intrinsic properties of the test mass or a constant biased external magnetic field (i.e. the Zeeman effect). Then the Stern-Gerlach interferometer can be created by a non-uniform external magnetic field. For example, the 1-dimensional Hamiltonian of the ideal gravimeter without noise based on the nitrogen-vacancy (NV) centre can be written as~\cite{PhysRevLett.125.023602}
\begin{equation}\label{original Hamiltonian}
    \hat{H} = \frac{\hat{p}^2}{2m} + g_{\rm NV}\mu_B\frac{\partial B}{\partial x}\hat{S}_z\hat{x} - \frac{\chi V}{2\mu_0}\left(\frac{\partial B}{\partial x}\right)^2\hat{x}^2 - mg\hat{x} + \hbar D\hat{S}_z^2.
\end{equation}
The term $\hbar D\hat{S}_z^2$ is the zero-field splitting of the NV centre with $D=(2\pi)2.8$\,GHz, which physically arises from the electron spin-spin dipole interaction~\cite{PhysRevB.90.235205}. Under this term, the spin of the test mass splits to eigenstates of $\hat{S}_z$, where $z$-direction is assumed as the direction of the NV-axis for simplicity, which can be changed to any axis in a real experiment. The spin-orbit coupling term $g_{\rm NV}\mu_B(\partial B/\partial x)\hat{S}_z\hat{x}$ arises the spin-magnetic field interaction, where $\mu_B$ is the Bohr magneton and $g_{\rm NV}$ is the Lande g-factor of NV centre. The quadratic term $\chi V/(2\mu_0) (\partial B/\partial x)^2\hat{x}^2$ is the diamagnetic energy of the diamond, with $\chi=-2.2\times10^{-5}$ and $V$ are the magnetic susceptibility and the volume of the diamond. The magnetic field gradient $\partial B/\partial x$ can be set as a time-independent constant~\cite{PhysRevLett.111.180403}, a piecewise function of time~\cite{PhysRevLett.125.023602}, several pulses~\cite{doi:10.1126/sciadv.abg2879}, or more complicated functions~\cite{PhysRevResearch.4.023087, Zhou:2024voj}. The term $mg\hat{x}$ accounts for the coupling to gravitational acceleration, which can be substituted with other coupling mechanisms for different experimental contexts. For instance, this sensor can also function as an accelerometer or seismometer by coupling to inertial forces~\cite{Zhao2022InertialMW, RevModPhys.89.035002}.

In order to create superpositions by the magnetic field, it is essential to avoid the procession of the spin such that spin states are pinned along the direction defined by the splitting term. For the Hamiltonian \eqref{original Hamiltonian}, the  $\hat{\mathbf{S}}\cdot\mathbf{B}$ term is supposed smaller than the zero-splitting term to avoid the Larmor precession. Since the energy of the zero-splitting term is $(2\pi)\hbar D\sim1.8\times10^{-24}\,\textrm{J}\sim g\mu_B\times10^{-1}\,\textrm{T}$, then the value of the magnetic field $|\mathbf{B}|$ is required smaller than 0.1\,T~\cite{PhysRevLett.130.113602}. 
In this case, the spin operator $\hat{S}_z$ can be treated as a c-number rather than an operator for ease of calculation. Then $s_z=\pm1$ of the interferometer corresponds to two paths, and each path of the interferometer is exactly a trajectory of a harmonic oscillator driven by external forces. Thus, the Hamiltonian for a single arm can be written as 
\begin{equation}\label{simplified Hamiltonian}
    \hat{H}(t) = \frac{\hat{p}^2}{2m} + \frac{1}{2}m\omega_0^2\hat{x}^2 - ma(t)\hat{x},
\end{equation}
where $ma(t)$ consists of the driven force $g_{\rm NV}\mu_B(\partial B/\partial x)s_z\hat{x}$ and the signal force $mg\hat{x}$, where the time-dependence of the Hamiltonian mainly comes from the time-dependent noise. The corresponding Schrodinger equation can be analytically solved by \emph{Lewis-Riesenfeld invariant} and \emph{quantum Arnold transformation}~\cite{Guerrero_2015, Aldaya_2011}, and the math details are summarized in Appendix\,\ref{appendix a}. Notably, other interferometers driven by spin-orbit coupling (SOC)~\cite{PhysRevA.109.063310} or SQUID~\cite{Johnsson2016} have Hamiltonians of similar forms, so the noise analysis in this paper is also valid for these types of interferometers.

\subsection{Time evolution of single path}

As is derived in Appendix\,\ref{appendix a}, the time-evolution operator for the Hamiltonian \eqref{simplified Hamiltonian} is
\begin{equation}\label{general solution}
    \hat{U}(t) = \mathrm{e}^{\frac{i}{\hbar}\int_0^tL[x(t'),\dot{x}(t')]\mathrm{d}t'}\hat{D}(\alpha_c(t))\mathrm{e}^{-\frac{i}{\hbar}\hat{H}_0t}\hat{D}(-\alpha_c(0)),
\end{equation}
where 
\begin{equation}
\begin{aligned}
    \hat{H}_0 &= \hat{p}^2/2m+m\omega_0^2\hat{x}^2/2, \\
    L[x(t),\dot{x}(t)] &= \frac{1}{2}m\dot{x}^2-\frac{1}{2}m\omega_0^2x+ma(t)x, \\
    \hat{D}(\alpha_c(t)) &= \exp\left[\alpha_c(t)\hat{a}^\dagger-\alpha_c^*(t)\hat{a}\right], \\
    \alpha_c(t) &= \sqrt{m\omega_0/2\hbar}(x_c(t)+ip_c(t)/m\omega_0). 
\end{aligned} 
\end{equation}
Here, $\hat{H}_0$ is the Hamiltonian of a simple harmonic oscillator without driven forces, $L[x(t),\dot{x}(t)]$ is the Lagrangian corresponding to the Hamiltonian \eqref{simplified Hamiltonian}, and $\hat{D}(\alpha_c(t))$ is the displacement operator defined by the complex coordinate $\alpha_c(t)$ of the instantaneous equilibrium position in the phase space, which satisfies the canonical equation in complex coordinate
\begin{equation}\label{dynamical alpha}
    \dot{\alpha}_c = -i\omega_0\alpha_c + i\sqrt{\frac{m}{2\hbar\omega_0}}a(t),
\end{equation}
of which the solution is
\begin{equation}\label{alpha t}
    \alpha_c(t) = \mathrm{e}^{-i\omega_0t}\left(\alpha_c(0)+i\sqrt{\frac{m}{2\hbar\omega_0}}\int_0^ta(t')\mathrm{e}^{i\omega_0t'}\mathrm{d}t'\right). 
\end{equation}

The physical interpretation of the time evolution operator $\hat{U}(t)$ in \eqref{general solution} is that the system is a simple oscillator without any driven force in the co-moving reference frame of the equilibrium position, where the displacement operators $\hat{D}(\alpha_c(t))$ and $\hat{D}(-\alpha_c(0))$ play the role of the reference translation in the classical phase space. Besides, $\hat{U}(t)$ also contains a path integral phase $\mathrm{e}^{\frac{i}{\hbar}\int_0^tL[x(t'),\dot{x}(t')]\mathrm{d}t'}$ along the classical trajectory of the equilibrium position $x_c(t)$ of the oscillator.

Consider an arbitrary initial state described by a density matrix $\hat{\rho}_0$ evolving as
\begin{equation}
    \hat{\rho}(t) = \hat{U}(t)\hat{\rho}_0\hat{U}^\dagger(t).
\end{equation}
From the point of view of Weyl quantization, the Wigner function is the Weyl-Wigner transform of the density matrix~\cite{Scully_Zubairy_1997}, i.e. 
\begin{equation}
    W(\alpha)=\int\text{Tr}[\hat{\rho}\hat{D}(\alpha')]\mathrm{e}^{\alpha'^*\alpha-\alpha'\alpha^*}\frac{\mathrm{d}^2\alpha'}{\pi^2},
\end{equation}
then the time-evolution of the Wigner function of an arbitrary state satisfies
\begin{equation}\label{Wigner time-evolution}
\begin{split}
    W(\alpha, t) &= W\left(\alpha_c(0)+\mathrm{e}^{i\omega_0t}(\alpha-\alpha_c(t)), t=0\right) \\
        &= W\left(\mathrm{e}^{i\omega_0t}\alpha-i\sqrt{\frac{m}{2\hbar\omega_0}}\int_0^ta(t')\mathrm{e}^{i\omega_0t'}\mathrm{d}t', t=0\right).
\end{split}
\end{equation}
Thus, the shape of the Wigner function of an arbitrary state doesn't change during time evolution, and its time evolution is only a transformation in the phase space composited by a translation $\alpha_c(t)$ plus a rotation with a constant angular speed $\omega_0$. Besides, the evolution of an arbitrary quantum state is completely determined by its classical trajectory $x_c(t)$ and momentum $p_c(t)$ satisfying the classical equations of motion \eqref{dynamical alpha} in the phase space.

Specially, if the interferometer is initially prepared as a coherent state $\ket{\beta_0}=\hat{D}(\beta_0)\ket{0}$, then it evolves as
\begin{equation}\label{coherent state time-evolution}
    \ket{\psi(t)} = \mathrm{e}^{i\phi(t)} \mathrm{e}^{-i\mathrm{Im}(\Delta\alpha_c\beta_0^*)} \mathrm{e}^{-\frac{1}{2}\omega_0 t} \ket{\alpha(t)},
\end{equation}
where 
\begin{align}
    \phi(t) &= \frac{1}{\hbar}\int_0^tL[x_c(t'), \dot{x}_c(t')]\,\mathrm{d}t', \\
    \alpha(t) &= \alpha_c(t) + \mathrm{e}^{-i\omega_0t}(\beta-\alpha_c(0)),
\end{align}
and the notation $\Delta\alpha_c\equiv\alpha_c(t)-\alpha_c(0)$. So every single arm remains in a coherent state during the time evolution, of which the wavefunction keeps in a Gaussian wave packet with a fixed width. Besides, the time evolution of an initial coherent state can be determined by the classical trajectory. Remarkably, there are additional phases $\mathrm{Im}(\Delta\alpha_c\beta_0^*)$ and $-\omega_0t/2$ related to the details of the initial quantum state.

\subsection{Time evolution of interferometer and witness in spin-space}

For an interferometer, the driven forces on its different arms are spin-dependent, so the corresponding time-evolution operators are different. In particular, the spatial part of the density matrix related to the spin element $\ket{i}\bra{j}$ evolves as $\hat{U}_i(t)\hat{\rho}_0\hat{U}^\dagger_j(t)$ with $i, j=\pm1$. Suppose the spin state is initially prepared as a pure state 
\begin{equation}
    \ket{S_z(t=0)} = \frac{\ket{1}+\ket{-1}}{\sqrt{2}},
\end{equation}
then the total density matrix of the interferometer at time $t$ is
\begin{equation}\label{total density matrix}
    \hat{\rho}_{\rm tot}(t) = \frac{1}{2}\sum\limits_{i,j=\pm1}\mathrm{e}^{i(\phi_i(t)-\phi_j(t))}\ket{i}\bra{j}\otimes\hat{\rho}_{ij}(t),
\end{equation}
where 
\begin{equation}
    \hat{\rho}_{ij}(t) \sim \hat{U}_i(t)\hat{\rho}_0\hat{U}^\dagger_j(t),
\end{equation}
and the phase $\phi_{i,j}(t)$ only contains the spatial path integral along the classical trajectories and the spin dynamics are ignored for simplicity.

At the final time $t_f$ of the experiment, the observable quantity is usually constructed by a witness operator $\hat{W}$ in the spin space, so the spatial wavefunctions have to be traced out, i.e.
\begin{equation}
    \hat{\rho}_{\rm spin}(t) = \mathrm{Tr}_{x}[\hat{\rho}(t)].
\end{equation} 
If the final position $\alpha_1(t)$ and $\alpha_{-1}(t)$ in the classical phase space are designed the same at the final time $t_f$, then all the spatial density matrix elements $\hat{\rho}_{ij}(t_f)$ are the same. As a consequence, the spin space and the spatial space are separable, i.e. $\hat{\rho}_{\rm tot}(t_f)=\hat{\rho}_{\rm spin}(t_f)\otimes\hat{\rho}(t_f)$, so the spin density matrix is simply
\begin{equation}\label{ideal rho spin}
    \hat{\rho}_{\rm spin}(t_f) = \frac{1}{2}\begin{pmatrix}
        1 & 0 & \mathrm{e}^{-i\phi_{\rm diff}} \\
        0 & 0 & 0 \\
        \mathrm{e}^{i\phi_{\rm diff}} & 0 & 1
    \end{pmatrix},
\end{equation}
with the notation $\phi_{\rm diff}\equiv\phi_{-1}(t_f)-\phi_1(t_f)$. 
A common-used witness to measure the differential phase $\phi_{\rm diff}$ is the \emph{Ramsey interferometry}~\cite{PhysRev.76.996, PhysRevLett.111.180403, 10.1063/1.5089550}. It is constructed by measuring the population of $\ket{S_z=0}$ state after a $\pi/2$-pulse operating on the state. As is derived in the Appendix\,\ref{appendix b}, the witness operator for Ramsey interferometry in spin-1 system is given by
\begin{equation}\label{Ramsey witness}
    \hat{W} = \begin{pmatrix}
        \frac{1}{2} & 0 & \frac{1}{2} \\
        0 & 0 & 0 \\
        \frac{1}{2} & 0 & \frac{1}{2}
    \end{pmatrix}.
\end{equation}

Then one can obtain the observable result for the Ramsey interferometry is
\begin{equation}
    \mathrm{Tr}[\hat{\rho}_{\rm spin}(t_f)\hat{W}] = \cos^2\frac{\phi_{\rm diff}}{2},
\end{equation}
which means that the population of the eigenstate $\ket{S_z=0}$ at the final time after a $\pi/2$-pulse is $\cos^2(\phi_{\rm diff}/2)$. For a Stern-Gerlach interferometer coupling to the local gravity by $\hat{H}_{\rm int}=mg\hat{x}$ in \eqref{original Hamiltonian}, the differential phase $\phi_{\rm diff}$ is proportional to $g$, so this device can be used as a gravimeter to measure the gravitational acceleration $g$ through the witness \eqref{Ramsey witness} of Ramsey interferometry~\cite{PhysRevLett.111.180403}.

\section{Dephasing Induced by Acceleration Noises}\label{section 3}

In a real experiment, there always exist some acceleration noises $\delta a(t)$ such as the vibration of the mechanical support of the system, the inertial force due to the vibration of the lab (known as the residual acceleration noise (RAN)~\cite{PhysRevResearch.3.023178, PhysRevD.107.104053}) and electric forces caused by ambient electric particles~\cite{PhysRevA.110.022412}. 

In most experimental situations, acceleration noises can be described by a (wide-sense) \emph{stationary Gaussian process} with a zero mean value. In particular, at any time $t$, the acceleration noise $\delta a(t)$ follows a Gaussian distribution. In addition, the \emph{wide-sense stationarity} requires the auto-correlation function of $\delta a(t)$ doesn't vary with respect to time $t$, i.e. $\mathbb{E}[\delta a(t)\delta a(t+\tau)]$ only depends on $\tau$, where $\mathbb{E}[\cdot]$ represents the statistical average of stochastic variables. According to the Wiener-Khinchin theorem, the auto-correlation function of $\delta a(t)$ is the Fourier transform of the corresponding \emph{power spectrum density} (PSD) 
\begin{equation}\label{auto-correlation of noise}
\begin{aligned}
    \mathbb{E}[\delta a(t)] &= 0, \quad \forall t, \\
    \mathbb{E}[\delta a(t)\delta a(t+\tau)] &= \int S_{aa}(\omega)\mathrm{e}^{-i\omega\tau}\mathrm{d}\omega, \quad \forall t,
\end{aligned}
\end{equation}
where $S_{aa}(\omega)=\lim\limits_{T\to\infty}|\delta a_T(\omega)|^2/T$ and $\delta a_T(\omega)$ is the Fourier transform of $\delta a(t)$ on the finite time domain $0\sim T$. Note that the second equation is independent with $t$ because the noise is assumed stationary. In addition, let $\tau=0$, then the auto-correlation function of $\delta a(t)$ gives the variance of $\delta a(t)$ as $\sigma_a^2(t)\equiv\mathbb{E}[(\delta a(t))^2] = \int S_{aa}(\omega)\mathrm{d}\omega$, which is also time-independent.

The acceleration noise $\delta a(t)$ couples to the test mass directly via the Hamiltonian $\hat{H}_{\rm noise}=-m\delta a(t)\hat{x}$, then the Hamiltonian \eqref{simplified Hamiltonian} for a single arm becomes to 
\begin{equation}\label{noisy Hamiltonian}
    \hat{H}(t) = \frac{\hat{p}^2}{2m} + \frac{1}{2}m\omega_0^2\hat{x}^2 - ma(t)\hat{x} - m\delta a(t)\hat{x},
\end{equation}
which is still of the form as \eqref{simplified Hamiltonian}. Hence, the time evolution operator \eqref{general solution} also holds when the noise is present, although the phase and the complex coordinate $\alpha_c(t)$ fluctuate. This section will concentrate on the dephasing, while the next section will disucss the decoherence of the spatial part of the test mass.

\subsection{Dephasing as a linear response to noise}

According to the time-evolution operator \eqref{general solution}, the spatial phase is precisely the path integral along the classical trajectories of the centre of mass. When the test mass experiences disturbances due to an acceleration noise, this phase is generally fluctuated by two leading-order channels. The first channel involves the path integral of the Lagrangian $m\delta a(t)x$ along the classical trajectory $x_c(t)$, while the second channel arises from the fluctuation of the classical trajectory $\delta x$ and velocity $\delta\dot{x}$ in the path integral of the ideal Lagrangian. The interplay between these two channels represents a higher-order effect and can be neglected. Thus, the phase fluctuation of a single arm can be expressed as
\begin{equation}
    \delta\phi = \frac{1}{\hbar}\int m\dot{x}\delta\dot{x} - m\omega_0^2x\delta x + ma(t)\delta x + m\delta a(t)x \mathrm{d}t.
\end{equation}
In order to evaluate the phase fluctuation $\delta\phi$, it is notable that the dynamical equation \eqref{dynamical alpha} can be written as another form as $\ddot{x} + \omega_0^2x = a(t)$, then one can find out an identity
\begin{equation}
    m\dot{x}\delta\dot{x} - m\omega_0^2x\delta x + ma(t)\delta x = m\frac{\mathrm{d}}{\mathrm{d}t}\left(\dot{x}\delta x\right).
\end{equation}
Thus, the phase fluctuation arising from the fluctuation of the trajectory $\delta x(t)$ and velocity $\delta\dot{x}(t)$ is a total derivative that will vanish after the path integral. Remarkably, some authors have already noticed that this channel of the phase fluctuation has to be taken into account in principle, but they neglected this effect by assuming it is a higher-order effect~\cite{PhysRevResearch.3.023178, PhysRevD.107.104053, PhysRevA.110.022412, PhysRevD.111.064004}. As is proved above, the dynamical equation ensures this effect is strictly negligible without any assumptions
\footnote{This conclusion can be generalized to a generic matter-wave interferometer under a general noise. Suppose the interferometer is built up through a Lagrangian $L[x_i(t), \dot{x}_i(t)]$ and the noise is described by a Lagrangian $\delta L[x_i(t), \dot{x}_i(t)]$. Denote the ideal trajectory as $x_i^0(t)$, and the ideal phase is $\phi^0=\int L[x_i^0, \dot{x}_i^0]/\hbar\mathrm{d}t$. Then the phase under noise at the leading order is
\begin{equation*}
\begin{split}
    \phi^0+\delta\phi &= \frac{1}{\hbar}\int L[x_i^0+\delta x_i, \dot{x}_i^0+\delta\dot{x}_i^0] + \delta L[x_i^0, \dot{x}_i^0]\mathrm{d}t \\ 
        &= \frac{1}{\hbar}\int L[x_i^0,\dot{x}_i^0] + \frac{\partial L}{\partial x_i}\bigg|_{(x_i^0,\dot{x}_i^0)}\delta x_i + \frac{\partial L}{\partial\dot{x}_i}\bigg|_{(x_i^0,\dot{x}_i^0)}\delta\dot{x}_i + \delta L[x_i^0, \dot{x}_i^0]\mathrm{d}t.
\end{split}
\end{equation*}
According to the Euler-Lagrangian equation $\partial L/\partial x_i=\mathrm{d}(\partial L/\partial\dot{x}_i)/\mathrm{d}t$, one can prove an identity
\begin{equation*}
    \frac{\partial L}{\partial x_i}\bigg|_{(x_i^0,\dot{x}_i^0)}\delta x_i + \frac{\partial L}{\partial\dot{x}_i}\bigg|_{(x_i^0,\dot{x}_i^0)}\delta\dot{x}_i = \frac{\mathrm{d}}{\mathrm{d}t}\left(\frac{\partial L}{\partial\dot{x}_i}\bigg|_{(x_i^0,\dot{x}_i^0)}\delta x_i\right).
\end{equation*}
Thus, the phase fluctuation arising from the fluctuation of the trajectory $\delta x_i$ and the velocity $\delta\dot{x}_i$ is the integral of a total derivative. Then one may read $\delta\phi=\int\delta L[x_i^0, \dot{x}_i^0]/\hbar\mathrm{d}t$ by neglecting this total derivative term. Note that the same conclusion has been proved by using Feynman's path integral method~\cite{Storey1994TheFP}.
}. Therefore, the phase fluctuation of a single path is simply
\begin{equation}\label{delta_phi}
    \delta\phi(t) = \frac{1}{\hbar}\int_0^t m\delta a(t')x(t')\mathrm{d}t'.
\end{equation}
So the phase fluctuation $\delta\phi(t)$ of a single path is an Ito integral of the stochastic noise $\delta a(t)$ and can be regarded as a linear response to $\delta a(t)$. Since the acceleration noise $\delta a(t)$ is assumed as a stationary Gaussian process, the phase fluctuation $\delta\phi(t)$ is a Gaussian process with a zero mean-value $\mathbb{E}[\delta\phi(t)]=0$ for arbitrary time $t$, although it is no longer stationary. The variance of $\delta\phi(t)$ can be computed as
\begin{equation}
    \sigma_\phi^2(t) \equiv \mathbb{E}[\left(\delta\phi(t)\right)^2] = \frac{m^2}{\hbar^2}\int S_{aa}(\omega)|x_t(\omega)|^2\mathrm{d}\omega,
\end{equation}
where $x_t(\omega)$ is the Fourier transform of the trajectory $x(t')$ in the finite time domain $0\sim t$, which can be written by a convolution between $x(\omega)$ and the sinc-function as $x_t(\omega)=x(\omega)\ast\text{sinc}(\omega t)$. Furthermore, $x(\omega)$ plays a role of the \emph{transfer function} (also known as \emph{frequency response function} or \emph{filter function}) of the linear response of $\delta\phi$ to $\delta a$~\cite{PhysRevD.99.104026, PhysRevD.107.104053, PhysRevD.111.064004, RevModPhys.89.035002}. 

For the interferometer with two paths, the phase fluctuations of both arms are linear responses to the acceleration noises, so the differential phase $\delta\phi_{\rm diff}$ is also a linear response to $\delta a$ and its variance is
\begin{equation}\label{dephasing factor and transfer function}
\begin{aligned}
    \delta\phi_{\rm diff} &= \frac{m}{\hbar}\int_0^{t_f}\delta a(t)(x_1(t)-x_{-1}(t))\mathrm{d}t, \\
    \sigma_{\phi_{\rm diff}}^2 &= \frac{m^2}{\hbar^2}\int S_{aa}(\omega)F(\omega)\mathrm{d}\omega,
\end{aligned}
\end{equation}
where 
\begin{equation}\label{transfer function}
    F(\omega) = |x_1(\omega)-x_{-1}(\omega)|^2,
\end{equation} 
is the transfer function or filter function of the linear response~\cite{PhysRevD.99.104026, PhysRevD.107.104053, PhysRevD.111.064004, RevModPhys.89.035002}.

\subsection{Properties of transfer function}

The transfer function has the following asymptotic behaviours. 
\begin{itemize}[leftmargin=8pt, noitemsep]
    \item $F(\omega)$ tends constant in the low-frequency limit. In particular, in the low-frequency limit $\omega\to0$, the factor $\mathrm{e}^{i\omega t}\to1$, then the Fourier transform satisfies $x_i(\omega)=\int x_i(t)\mathrm{e}^{i\omega t}\mathrm{d}t\approx\int x_i(t)\mathrm{d}t$ which is independent with $\omega$, so $F(\omega)$ is approximately constant to $\omega$ in the low-frequency limit.
    \item $F(\omega)$ decreases as $\omega^{-2k}$ for some index $k$ in the high-frequency limit. In particular, in the high-frequency limit, the trajectories can be expanded as Taylor's series as $x_1(t)-x_{-1}(t)=\sum_{n=k}^\infty c_nt^n$ with the leading order as $t^k$, then the corresponding Fourier transform is $x_1(\omega)-x_{-1}(\omega)=\sum_{n=k}^{\infty}c_n\omega^{-n-1}$, so the transfer function is proportional to $\omega^{-2k-2}$ at the leading order in the high-frequency limit.
\end{itemize}

According to the dynamical equation $\ddot{x}_i+\omega_0^2=a_i(t)$ which is associated with the Hamiltonian \eqref{noisy Hamiltonian}, one can obtain an equation $(-\omega^2+\omega_0^2)x_i(\omega)=a_i(\omega)$ in frequency space. Hence, the transfer function can be written as
\begin{equation}
    F(\omega) = \frac{|a_1(\omega)-a_{-1}(\omega)|^2}{(\omega^2-\omega_0^2)^2}.
\end{equation}
If $a_i(\omega)$ has no singularity, the transfer function is a meromorphic function with two second-order poles $\pm\omega_0$. Since the $S_{aa}(\omega)$ and $a_i(\omega)$ are symmetric in the frequency domain, the variance of the phase fluctuation can be computed by the residue theorem as
\begin{equation}
\begin{split}
    \sigma_{\phi_{\rm diff}}^2 &= 4\pi\frac{m^2}{\hbar^2}\frac{\mathrm{d}}{\mathrm{d}\omega}\left[\frac{S_{aa}(\omega)}{(\omega+\omega_0)^2}\left|a_1(\omega)-a_{-1}(\omega)\right|^2\right]\bigg|_{\omega=\omega_0} \\
        &= \pi\frac{m^2}{\hbar^2\omega_0^3}S_{aa}(\omega_0)|a_1(\omega_0)-a_{-1}(\omega_0)|^2,
\end{split}
\end{equation}
where the math properties of $a_i(\omega)$ and $S_{aa}(\omega)$ are assumed good enough such that their derivatives don't have dominant contributions to $\sigma_{\phi_{\rm diff}}^2$. In this sense, the transfer function $F(\omega)$ is approximately a delta function, i.e. 
\begin{equation}
    F(\omega)\approx A\delta(\omega-\omega_0),
\end{equation}
with a factor $A=\pi|a_1(\omega_0)-a_{-1}(\omega_0)|^2/\omega_0^3$.
\begin{figure*}
    \centering
    \begin{subfigure}{0.45\textwidth}
        \includegraphics[scale=0.5]{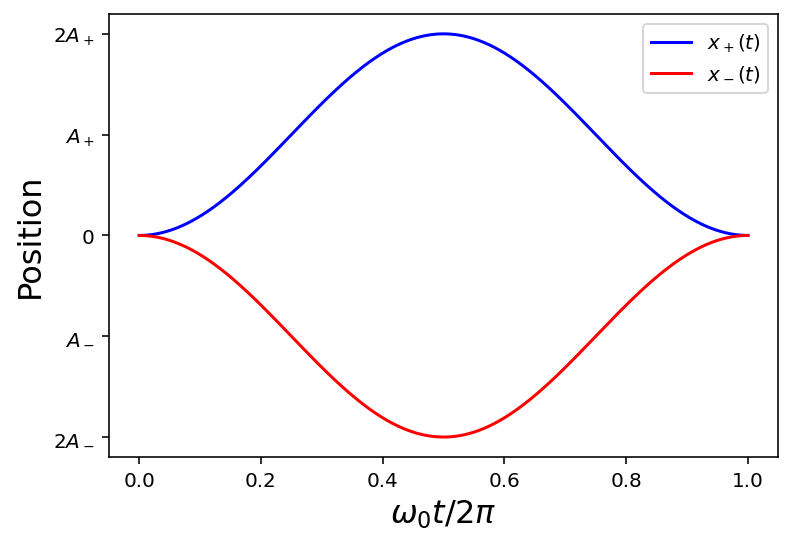}
        \subcaption{}
    \end{subfigure}
    \hfill
    \begin{subfigure}{0.45\textwidth}
        \includegraphics[scale=0.5]{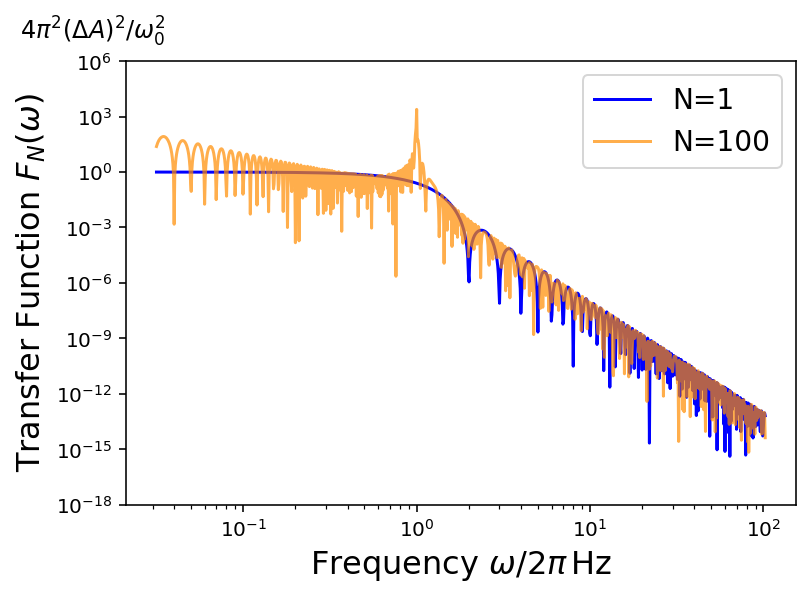}
        \subcaption{}
    \end{subfigure}
    \caption{\small (a) The unperturbed trajectories of an interferometer under a time-independent magnetic field $B=\eta x$. (b) The corresponding transfer function for the trajectories with $N$ periods.}
    \label{trajectories and transfer function}
\end{figure*}

For example, consider a time-independent magnetic field $B(t)=\eta x$ with $\eta\equiv\partial B/\partial x$. As is shown in Fig.\,\ref{trajectories and transfer function} (a), the unperturbed trajectories are 
\begin{equation}\label{trajectories for constant magnetic gradient}
    x_{\pm}(t)=A_{\pm}(1-\cos\omega_0 t),\quad \text{with } A_{\pm}=\pm \frac{g_{\rm NV}\mu_B\eta}{m\omega_0^2}.
\end{equation}
where the gravitational forces are ignored for simplicity. If the total experiment time is $2\pi N/\omega_0$, i.e. the trajectories repeat $N$ periods in a single run of the experiment, then the corresponding transfer function is
\begin{equation}\label{N transfer function}
    F_N(\omega) = \frac{4(\Delta A)^2\omega_0^4\sin^2\frac{N\pi\omega}{\omega_0}}{\omega^2(\omega^2-\omega_0^2)^2},
\end{equation}
where $\Delta A\equiv A_+-A_-$. According to \eqref{N transfer function}, $F_N(\omega)$ tends to a constant $4\pi^2N^2(\Delta A)^2/\omega_0^2$ in the low-frequency limit $\omega\to0$ and decreases with respect to $\omega^{-6}$ in the high-frequency limit $\omega\to\infty$. As is shown in Fig.\,\ref{trajectories and transfer function} (b), the transfer function has a peak at $\omega=\omega_0$, so it is approximated as a delta function $\delta(\omega-\omega_0)$, except when the number of periods $N=1$, in which case this peak is hidden. 
In fact, as derived in Appendix \ref{appendix c}, the variance of the phase fluctuation in this system can be evaluated using the residue theorem, yielding
\begin{equation}\label{sigma phi}
    \sigma_{\phi_{\rm diff}}^2 = 4\pi m^2\frac{N^2(\Delta A)^2S_{aa}(\omega_0)}{\omega_0\hbar^2}.
\end{equation}

\section{Position Localisation Decoherence and Humpty-Dumpty Problem}\label{section 4}

As is pointed out at the beginning of the previous section, the time-evolution operator of the interferometer under the noise $\delta a(t)$ has the same form as the ideal situation. Thus, the time evolution \eqref{Wigner time-evolution} of the Wigner function of an arbitrary initial state becomes
\begin{equation}\label{noisy Wigner}
    W(\alpha, t) = W(\alpha_c(0)+\mathrm{e}^{i\omega_0t}(\alpha-\alpha_c(t)-\delta\alpha(t)), t=0).
\end{equation}

The first notable conclusion is that the acceleration noise $\delta a(t)$ doesn't change the shape of the Wigner function of an arbitrary initial state, and it only affects the path integral phase $\delta\phi$ and the classical trajectory $\delta\alpha$ in phase space.

\subsection{Fluctuation of classical phase space $\delta\alpha$ of single arm}

For a coherent state given by \eqref{coherent state time-evolution}, the wavefunction under the acceleration noise is still a coherent state, although its parameter $\beta(t)=\alpha_c(t)+\mathrm{e}^{-i\omega_0 t}(\beta_0-\alpha_c(0))$ changes to $\beta'(t)=\beta(t)+\delta\alpha(t)$.
According to the time-evolution \eqref{alpha t} of the complex coordinate $\alpha_c$ in the phase space, its fluctuation $\delta\alpha$ under the acceleration noise $\delta a(t)$ is given by
\begin{equation}\label{delta alpha}
    \delta\alpha(t) = i\sqrt{\frac{m}{2\hbar\omega_0}}\mathrm{e}^{i\omega_0t}\int_0^t\delta a(t')\mathrm{e}^{-i\omega_0t'}\mathrm{d}t'.
\end{equation}
So $\delta\alpha(t)$ is also a linear response to the acceleration noise $\delta a(t)$. Besides, $\delta\alpha(t)$ is also a Gaussian process with a zero mean-value $\mathbb{E}[\delta\alpha(t)]=0$ and its auto-correlation function is
\begin{equation}\label{auto-correlation of alpha}
\begin{split}
    \mathbb{E}[\delta\alpha(t)\delta\alpha^*(t+\tau)] &= \frac{m}{2\hbar\omega_0}\mathrm{e}^{-\frac{i}{2}\omega_0\tau} t(t+\tau) \int S_{aa}(\omega)\mathrm{e}^{-\frac{i}{2}\omega\tau} \\
        \times& \text{sinc}\frac{(\omega-\omega_0)t}{2}\text{sinc}\frac{(\omega-\omega_0)(t+\tau)}{2}\mathrm{d}\omega,
\end{split}
\end{equation}
Remarkably, $\mathbb{E}[\delta\alpha(t)\delta\alpha^*(t+\tau)]$ relies on both $t$ and $\tau$, so $\delta\alpha$ is not stationary. Hence, the Wiener-Khinchin theorem doesn't hold for $\delta\alpha$. 

In particular, consider the dynamical equation \eqref{dynamical alpha} of $\alpha_c$,  one may obtain a frequency response relation as 
\begin{equation}
\begin{aligned}
    \delta\alpha(\omega) = \chi(\omega)\delta a(\omega),\quad \text{with }
    \chi(\omega) = \frac{\sqrt{m/(2\hbar\omega_0)}}{\omega-\omega_0}.
\end{aligned}
\end{equation}
Here, the factor $\chi(\omega)$ is a generalized \emph{mechanical susceptibility} of $\alpha_c(\omega)$. Then the PSD of $\delta\alpha$ is
\begin{equation}
    S_{\alpha\alpha}(\omega) = \frac{m}{2\hbar\omega_0}\frac{S_{aa}(\omega)}{(\omega-\omega_0)^2}.
\end{equation}
Therefore, one may check that the relationship between the auto-correlation function and PSD of $\delta\alpha$ is
\begin{equation}
\begin{split}
    \mathbb{E}[\delta\alpha(t)\delta\alpha^*(t+\tau)] &= \int S_{\alpha\alpha}(\omega)\mathrm{e}^{-i\omega\tau} \bigg(1+\mathrm{e}^{-i(\omega-\omega_0)\tau} \\
        &-\mathrm{e}^{i(\omega-\omega_0)t}-\mathrm{e}^{-i(\omega-\omega_0)(t+\tau)}\bigg)\mathrm{d}\omega.
\end{split}
\end{equation}

In the auto-correlation function \eqref{auto-correlation of alpha} of $\delta\alpha$, let $\tau=0$, then one may have the variance of $\delta\alpha$ at any time $t$ as
\begin{equation}\label{sigma alpha t}
    \sigma_\alpha^2(t) \equiv \mathbb{E}[|\delta\alpha(t)|^2] = \frac{m}{2\hbar\omega_0}t^2\int S_{aa}(\omega)\text{sinc}^2\frac{(\omega-\omega_0)t}{2}\mathrm{d}\omega.
\end{equation}
Thus, $\delta\alpha$ can be regarded as a generalized diffusion process. In the short-time and long-time limits, its variance can be approximated as \footnote{Here, asymptotic behaviours of the sinc-function have been used. That is, $\text{sinc}^2((\omega-\omega_0)t/2)=1+\mathcal{O}(t^2)$ for small $t$, and $\text{sinc}^2((\omega-\omega_0)t/2)=2\pi\delta(\omega-\omega_0)/t$ for large $t$.}
\begin{equation}
    \sigma_\alpha^2(t) \approx \left\{ \begin{aligned}
        D_1t^2,\quad t\ll 1, \\
        D_2t,\quad t\to\infty,
    \end{aligned} \right.
\end{equation}
where the generalized diffusion coefficients are given by $D_1=m\sigma_a^2/(2\hbar\omega_0)$ and $D_2=\pi mS_{aa}(\omega_0)/(\hbar\omega_0)$. Notably, in the long-time limit, $\delta\alpha$ exhibits behaviour analogous to standard Brownian motion. That is, the diffusion process becomes \emph{Markovian} or \emph{memoryless} as $t\to\infty$, implying that its long-time diffusion is independent of its initial diffusion history.  

Specially, at the final time $t_f=2N\pi/\omega_0$, the variance is 
\begin{equation}\label{sigma2_alpha}
    \sigma_\alpha^2(t_f) = \frac{2\pi^2N^2m}{\hbar\omega_0^3}\int S_{aa}(\omega)\frac{\sin^2(N\pi\omega/\omega_0)}{N^2\pi^2(\omega/\omega_0-1)^2}\mathrm{d}\omega,
\end{equation}
which can be evaluated by the residue theorem, as shown in Appendix \ref{appendix c}. This yields
\begin{equation}\label{sigma alpha tf}
    \sigma_\alpha^2(t_f) = \frac{2\pi^2Nm}{\hbar\omega_0^2}S_{aa}(\omega_0).
\end{equation}
Physically, it can be interpreted that the test mass only resonants with the noise at frequency $\omega_0$, which is the intrinsic frequency of the interferometer. Remarkably, this result coincides with the long-time limit expression $\sigma_{\alpha}^2(t)=D_2t$, although $t_f$ does not necessarily fall within the asymptotic regime.

\subsection{Fidelity loss of a single arm}

Since the trajectory of a single arm is disturbed by the noise, the corresponding quantum state is also affected, which is usually characterised by the \emph{Uhlmann fidelity}~\cite{UHLMANN1976273, PhysRevX.14.021022}. In particular, denote the ideal and noisy density matrices of the single arm at an arbitrary time $t$ as
\begin{equation}
    \hat{\rho}_{\rm id}(t)\equiv\hat{U}_{\rm id}(t)\hat{\rho}_0\hat{U}_{\rm id}^\dagger(t),\quad \hat{\rho}_{\rm n}(t)\equiv\hat{U}_{\rm n}(t)\hat{\rho}_0\hat{U}_{\rm n}^\dagger(t),
\end{equation}
where $\hat{U}_{\rm id}(t)$ and $\hat{U}_{\rm n}(t)$ are given by \eqref{general solution} with complex coordinates $\alpha_c(t)$ and $\alpha_c(t)+\delta\alpha(t)$. Then the corresponding fidelity is defined as
\begin{equation}
    F(\hat{\rho}_{\rm id}(t), \hat{\rho}_{\rm n}(t)) \equiv \text{Tr}\left[\sqrt{\sqrt{\hat{\rho}_{\rm id}(t)} \hat{\rho}_{\rm n}(t) \sqrt{\hat{\rho}_{\rm id}(t)}}\right].
\end{equation} 
It is not difficult to prove that~\footnote{Here, the following identities are used: 
\begin{equation*}
\begin{aligned}
    \sqrt{\hat{\rho}_{\rm id}(t)} &= \hat{U}_{\rm id}(t)\sqrt{\hat{\rho}_0}\hat{U}_{\rm id}^\dagger(t), \\ 
    \hat{U}_{\rm id}^\dagger(t)\hat{U}_{\rm n}(t) &= \hat{D}\left(\delta\alpha(t)\mathrm{e}^{i\omega_0t}\right)\mathrm{e}^{i\delta\phi(t)}.
\end{aligned}
\end{equation*}
}
\begin{equation}
    F(\hat{\rho}_{\rm id}(t), \hat{\rho}_{\rm n}(t)) = F\left(\hat{\rho}_0, \hat{D}(\delta\alpha(t)\mathrm{e}^{i\omega_0t})\hat{\rho}_0\hat{D}(-\delta\alpha(t)\mathrm{e}^{i\omega_0t})\right).
\end{equation}
So the fidelity between $\hat{\rho}_{\rm id}(t)$ and $\hat{\rho}_{\rm n}(t)$ is exactly the fidelity of the initial state $\hat{\rho}_0$ and its fluctuation under $\hat{D}(\delta\alpha(t)\mathrm{e}^{i\omega_0t})$.
Specially, if the system is initially prepared as a pure state $\hat{\rho}_0=\ket{\psi_0}\bra{\psi_0}$, then the fidelity is simply the contrast 
\begin{equation}
    F(\hat{\rho}_{\rm id}(t), \hat{\rho}_{\rm n}(t)) = |\bra{\psi_0}\hat{D}(\delta\alpha(t)\mathrm{e}^{i\omega_0t})\ket{\psi_0}|.
\end{equation}
In addition, if the system is initially prepared as a coherent state, then the fidelity is
\begin{equation}
    F(\hat{\rho}_{\rm id}(t), \hat{\rho}_{\rm n}(t)) = \exp\left[-\frac{|\delta\alpha(t)|^2}{2}\right],
\end{equation}
which is exactly the overlap of two Gaussian wave-packets.

\subsection{Position localisation decoherence}

It is notable that the randomness of the trajectory fluctuation $\delta\alpha(t)$ will cause a \emph{position localisation decoherence}~\cite{PhysRevA.84.052121} on the single arm of the interferometer~\cite{Pino_2018, PhysRevLett.132.023601}. For an ideal interferometer, the matrix element of the density matrix $\hat{\rho}_{\rm id}(t)=\hat{U}_{\rm id}(t)\hat{\rho}_0\hat{U}_{\rm id}^\dagger(t)$ satisfies 
\begin{equation}
\begin{aligned}
    &\bra{x_1}\hat{\rho}_{\rm id}(t)\ket{x_2} = \bra{x'_1(t)}\hat{\rho}_0\ket{x'_2(t)}, \\
    &x'_i(t) = \left(x_i-\sqrt{\frac{2\hbar}{m\omega_0}}\text{Re}\alpha(t)\right)\cos\omega_0t + \sqrt{\frac{2\hbar}{m\omega_0}}\text{Re}\alpha(0).
\end{aligned}
\end{equation}
So the time-evolution of the density matrix element is simply the oscillation defined by the Hamiltonian \eqref{simplified Hamiltonian} without noise, which has been concluded by the Wigner function \eqref{Wigner time-evolution}. Besides, one can prove that the \emph{purity} of the density matrix $\mathcal{P}_{\rm id}(t)\equiv\text{Tr}\left[\hat{\rho}_{\rm id}^2(t)\right]$ keeps invariant according to the following calculation
\begin{equation}
\begin{split}
    \mathcal{P}_{\rm id}(t) &= \int\mathrm{d}x_1\mathrm{d}x_2\bra{x_1}\hat{\rho}_{\rm id}(t)\ket{x_2}\bra{x_2}\hat{\rho}_{\rm id}(t)\ket{x_1} \\
        &= \int\mathrm{d}x'_1(t)\mathrm{d}x'_2(t)\bra{x'_1(t)}\hat{\rho}_0\ket{x'_2(t)}\bra{x'_2(t)}\hat{\rho}_0\ket{x'_1(t)} \\
        &= \text{Tr}\left[\hat{\rho}_0^2\right] \equiv \mathcal{P}_0. 
\end{split}
\end{equation}

On the other hand, for the interferometer under noises, according to the identity $\hat{U}_{\rm n}(t)=\hat{D}(\delta\alpha(t))\hat{U}_{\rm id}(t)\mathrm{e}^{i\delta\phi(t)}$ with some phase $\delta\phi(t)$, the matrix element of the density matrix $\hat{\rho}_{\rm n}(t)=\hat{U}_{\rm n}(t)\hat{\rho}_0\hat{U}_{\rm n}^\dagger(t)$ is 
\begin{equation}
    \bra{x_1}\hat{\rho}_{\rm n}(t)\ket{x_2} = \bra{x'_1(t)}\hat{D}(\delta\alpha(t))\hat{\rho}_0\hat{D}^\dagger(\delta\alpha(t))\ket{x'_2(t)}.
\end{equation}
The ensemble average of this matrix element can be computed as
~\footnote{To compute the ensemble average of $\bra{x_1}\hat{\rho}_{\rm n}(t)\ket{x_2}$, one can consider the Glauber's P-representation~\cite{Scully_Zubairy_1997} of the initial density matrix 
\begin{equation*}
    \hat{\rho}_0=\int P(\beta)\ket{\beta}\bra{\beta}\mathrm{d}^2\beta.
\end{equation*}
According to the identity
\begin{equation*}
    \hat{D}(\delta\alpha(t))\ket{\beta}=\mathrm{e}^{i\phi(\beta,\delta\alpha)}\ket{\beta+\delta\alpha(t)},
\end{equation*}
the matrix element of $\hat{\rho}_{\rm n}(t)$ becomes to 
\begin{equation*}
    \bra{x_1}\hat{\rho}_{\rm n}(t)\ket{x_2} = \int P(\beta)\braket{x'_1(t)|\beta+\delta\alpha(t)}\braket{\beta+\delta\alpha(t)|x'_2(t)}\mathrm{d}^2\beta.
\end{equation*}
Based on the wavefunction of a coherent state, one can obtain the ensemble average of the matrix element in the P-representation satisfying
\begin{equation*}
\begin{split}
    &\mathbb{E}\left[\braket{x'_1(t)|\beta+\delta\alpha(t)}\braket{\beta+\delta\alpha(t)|x'_2(t)}\right] \\
    =& \braket{x'_1(t)|\beta}\braket{\beta|x'_2(t)}\mathrm{e}^{-\frac{m\omega_0}{\hbar}\sigma_\alpha^2(t)(x'_1(t)-x'_2(t))^2}.
\end{split}
\end{equation*}
Then one can obtain the result \eqref{rho_xy decoherence}.
}
\begin{equation}\label{rho_xy decoherence}
    \bra{x_1}\mathbb{E}\left[\hat{\rho}_{\rm n}(t)\right]\ket{x_2} = \bra{x'_1(t)}\hat{\rho}_0\ket{x'_2(t)}\mathrm{e}^{-\frac{m\omega_0}{\hbar}\sigma_\alpha^2(t)(x'_1(t)-x'_2(t))^2}.
\end{equation}
Consequently, in the long-time limit, off-diagonal terms of $\hat{\rho}_{\rm n}(t)$ decays exponentially and only the diagonal terms of $\hat{\rho}_{\rm n}(t)$ keeps non-zero, so the corresponding purity approaches to zero.
Furthermore, since the variance of $\delta\alpha$ is approximately proportional to $t$ in the long time limit, then there is a well-defined decoherence rate
\begin{equation}
    \gamma = \frac{\pi m^2}{\hbar^2}(x_1-x_2)^2S_{aa}(\omega_0).
\end{equation}
It is noteworthy that a decoherence rate proportional to $(x_1-x_2)^2$ is related to the following master equation~\cite{PhysRevA.84.052121, Pino_2018}
\begin{equation}\label{master equation}
    \frac{\mathrm{d}}{\mathrm{d}t}\hat{\rho} = \frac{1}{i\hbar}[\hat{H}, \hat{\rho}] - \Lambda[\hat{x}, [\hat{x}, \hat{\rho}]],
\end{equation}
where $\Lambda=\pi m^2S_{aa}(\omega_0)/\hbar^2$, which is the same as~\cite{Pino_2018} up to a factor $2\pi$. This master equation can be derived directly from the Hamiltonian \eqref{simplified Hamiltonian}, and the math details are summarized in \ref{appendix c}. 
It is noteworthy that the decoherence due to the environmental scattering in the long-wavelength limit also follows the same master equation~\cite{schlosshauer2007quantum, PhysRevA.84.052121}, so the scattering of long-wavelength particles like photons can be regarded as a classical stochastic noise in a long-time limit.

\subsection{Loss of witness}

\begin{figure*}
    \begin{subfigure}{0.25\textwidth}
        \begin{tikzpicture}[scale=0.6]
            \filldraw[red] (0,0) circle (0.1);
            \filldraw[red] (6.28,0) circle (0.1);
            \draw[domain=0:6.28,smooth] plot (\x, {sin(\x/2 r)});
    
            \filldraw[red] (6.28, 1.256) circle (0.1);
            \draw[domain=0:6.28,smooth] plot (\x, {sin(\x/2 r)+\x/5 +sin(2*\x r)/6+sin(6*\x r)/15});
        \end{tikzpicture}
        \subcaption{}
    \end{subfigure}
    \hfill
    \begin{subfigure}{0.35\textwidth}
        \begin{tikzpicture}[scale=0.8]
            \filldraw[red] (0,0) circle (0.1);
            \filldraw[red] (6.28,0) circle (0.1);
            \draw[domain=0:6.28,smooth] plot (\x, {sin(\x/2 r)});
            \draw[domain=0:6.28,smooth] plot (\x, {-sin(\x/2 r)});
    
            \filldraw[red] (6.28, 1.256) circle (0.1);
            \draw[domain=0:6.28,smooth] plot (\x, {sin(\x/2 r)+\x/5 +sin(2*\x r)/6+sin(6*\x r)/15});
            \draw[domain=0:6.28,smooth] plot (\x, {-sin(\x/2 r)+\x/5 +sin(2*\x r)/6+sin(6*\x r)/15});
        \end{tikzpicture}
        \subcaption{}
    \end{subfigure}
    \hfill
    \begin{subfigure}{0.35\textwidth}
        \begin{tikzpicture}[scale=0.8]
            \filldraw[red] (0,0) circle (0.1);
            \filldraw[red] (6.28,0) circle (0.1);
            \draw[domain=0:6.28,smooth] plot (\x, {sin(\x/2 r)});
            \draw[domain=0:6.28,smooth] plot (\x, {-sin(\x/2 r)});
    
            \filldraw[red] (6.28, 0.628) circle (0.1);
            \filldraw[red] (6.28, -0.628) circle (0.1);
            \draw[domain=0:6.28,smooth] plot (\x, {sin(\x/2 r)+\x/10 +sin(2*\x r)/6+sin(6*\x r)/15});
            \draw[domain=0:6.28,smooth] plot (\x, {-sin(\x/2 r)-\x/10 +sin(2*\x r)/6+sin(6*\x r)/15});
        \end{tikzpicture}
        \subcaption{}
    \end{subfigure}
    \caption{\small The trajectories affected by acceleration noises, where (a) for a single path, (b) for acceleration noises and (c) for path-dependent noises. As is shown in (b), under common mode noises, the final points in phase space of both arms translate the same, so the two paths can be overlapped properly. As is shown in (c), under path-dependent noises, the points of both arms translate differently, so there is a Loschmidt echo loss.}
    \label{Fig. uniform vs non-uniform}
\end{figure*}

For an interferometer with two arms, the trajectory fluctuations $\delta\alpha(t)$ of both paths are the same according to \eqref{delta alpha} as long as both arms couples to the noise in the same way. As is proved by \eqref{noisy Wigner}, the Wigner functions of the arms are totally determined by their classical trajectories in phase space, so the two arms can meet with each other properly at the final time, that is, the contrast of the two arms keeps 1, although the fidelity of each arm gets lost, shown as Fig.\ref{Fig. uniform vs non-uniform} (b). In addition, the position localisation decoherence also has no influence on decoherence in spin space, because $\hat{\rho}_{\rm spin}$ only relies on the diagonal terms of $\hat{\rho}_{\rm n}(t)$ which don't decay during evolution according to \eqref{rho_xy decoherence}. 

Thus, the spin space and the coordinate space are still separable, and the noise only contributes a fluctuation of the differential phase $\delta\phi_{\rm diff}$ to the ideal density matrix \eqref{ideal rho spin} in the spin space. It can be regarded as a type of common mode noise cancellation, which has been applied in atomic interferometers (AI)~\cite{PhysRevLett.81.971, PhysRevA.89.023607, WANG201634}. Notably, this effect is usually used to reduce dephasing in AIs, while it mitigates position localisation decoherence and fidelity loss in SGIs. 

As discussed in sub-section 3.1, the phase fluctuations of both arms follow Gaussian distributions, so the ensemble average of the random phase factor contributes an exponential decay factor $\mathbb{E}[\mathrm{e}^{i\delta\phi_{\rm diff}}]=\mathrm{e}^{-\sigma_{\rm diff}^2/2}$. Then the density matrix in spin space under the ensemble average is
\begin{equation}\label{noisy rho spin}
    \mathbb{E}[\hat{\rho}_{\rm spin}(t_f)] = \frac{1}{2}\begin{pmatrix}
        1 & 0 & \mathrm{e}^{-\frac{\sigma_{\phi_{\rm diff}}^2}{2}}\mathrm{e}^{-i\phi_{\rm diff}} \\
        0 & 0 & 0 \\
        \mathrm{e}^{-\frac{\sigma_{\phi_{\rm diff}}^2}{2}}\mathrm{e}^{i\phi_{\rm diff}} & 0 & 1
    \end{pmatrix}.
\end{equation}
Recall that ideal value of the witness of the Ramsey interferometry is $\mathrm{Tr}[\hat{\rho}_{\rm spin}(t_f)\hat{W}] = \cos^2(\phi_{\rm diff}/2)$. For the spin density matrix \eqref{noisy rho spin} under noise, the observable result of the witness is
\begin{equation}
    \braket{W} \equiv \mathrm{Tr}\left[\mathbb{E}[\hat{\rho}_{\rm spin}(t_f)]\hat{W}\right] = \frac{1}{2}\left[1+\mathrm{e}^{-\frac{\sigma_{\phi_{\rm diff}}^2}{2}}\cos\phi_{\rm diff}\right],
\end{equation}
and the corresponding loss of witness is
\begin{equation}
    \Delta W = \frac{\mathrm{e}^{-\frac{\sigma_{\phi_{\rm diff}}^2}{2}}-1}{2}\cos\phi_{\rm diff} \approx -\frac{\sigma_{\phi_{\rm diff}}^2}{2}\cos\phi_{\rm diff}.
\end{equation}
Therefore, the loss of witness is proportional to the variance of the phase fluctuation. Recall that $\sigma^2_{\phi_{\rm diff}}$ is approximately proportional to $S_{aa}(\omega_0)$, so the loss of witness $\Delta W$ should be mitigated by controlling the PSD at the intrinsic frequency $\omega_0$ of the interferometer in a real experiment.

\section{Leading-order Effects of Higher Order Noises}\label{section 5}

Apart from acceleration noises, there exist noises coupling to the interferometer in other ways such as $m\delta\omega^2(t)\hat{x}^2$ or $\mu\delta\eta(t)\hat{S}_z\hat{x}$ with $\eta(t)\equiv\partial B(t)/\partial x$. In order to study these noises analytically, it is necessary in principle to solve the corresponding stochastic Schrodinger equation for these coupling Hamiltonians, which can be a great challenge in math. 
However, by treating some operators as c-numbers, these noises can be regarded as path-dependent acceleration noises. Then the leading order effects of these noises can be studied through the analytic solution for acceleration noises, although some higher-order properties can be lost. 

As is shown in Fig.\ref{Fig. uniform vs non-uniform} (c), the final states of the two paths are usually not able to meet with each other, i.e. $\alpha_1(t_f)\neq\alpha_{-1}(t_f)$, so the total density matrix is not separable. 
As a consequence, apart from the dephasing effect characterised by a decay factor $\mathrm{e}^{-\sigma_{\phi_{\rm diff}}^2/2}$, the off-diagonal terms of the spin density matrix have an additional decay factor arising from the partial trace of the spatial degrees of freedom. Therefore, the spin density matrix under path-dependent noises is given by 
\begin{equation}
    \mathbb{E}[\hat{\rho}_{\rm spin}(t_f)] = \frac{1}{2}\begin{pmatrix}
        1 & 0 & \mathbb{E}\left[C\right]\mathrm{e}^{-\frac{\sigma_{\phi_{\rm diff}}^2}{2}}\mathrm{e}^{-i\phi_{\rm diff}} \\
        0 & 0 & 0 \\
        \mathbb{E}\left[C\right]\mathrm{e}^{-\frac{\sigma_{\phi_{\rm diff}}^2}{2}}\mathrm{e}^{i\phi_{\rm diff}} & 0 & 1
    \end{pmatrix},
\end{equation}
where
\begin{equation}\label{fidelity between two paths}
    C \equiv \text{Tr}\left[\sqrt{\sqrt{\hat{\rho}_1(t_f)} \hat{\rho}_{-1}(t_f) \sqrt{\hat{\rho}_1(t_f)}}\right],
\end{equation}
is the fidelity between the two paths at the final time, which is the generalisation of the contrast $C\equiv\braket{\psi_1(t_f)|\psi_{-1}(t_f)}$ of pure states to mixed states.

There are two decay factors $\mathrm{e}^{-\sigma_{\phi_{\rm diff}}^2/2}$ and $\mathbb{E}\left[C\right]$ in the off-diagonal terms of the density matrix. The first factor characterises the dephasing effect, while the second term corresponds to the combination of the position localisation decoherence and the Humpty-Dumpty problem.

\subsubsection*{Dephasing}

For the dephasing effect, the fluctuation of the differential phase of the interferometer is given by 
\begin{equation}
    \delta\phi_{\rm diff} = \frac{m}{\hbar}\int_0^{t_f}(\delta a_1(t)x_1(t)-\delta a_{-1}(t)x_{-1}(t))\mathrm{d}t.
\end{equation}
In this case, the variance of the phase fluctuation is determined by the correlation between $\delta a_1(t)$ and $\delta a_{-1}(t)$, which is characterised by their cross-PSD $S_{a_1a_{-1}}$. Hence, the variance $\sigma_{\phi_{\rm diff}}$ can be formulated as
\begin{equation}\label{higher noise sigma_phi}
\begin{split}
    \sigma_{\phi_{\rm diff}}^2 =& \frac{m^2}{\hbar^2}\int S_{a_1a_1}(\omega)F_{11}(\omega) + S_{a_{-1}a_{-1}}(\omega)F_{-1-1}(\omega) \\
        &+ S_{a_1a_{-1}}(\omega)F_{1,-1}(\omega) + S_{a_{-1}a_1}(\omega)F_{-1,1}(\omega)\mathrm{d}\omega,
\end{split}
\end{equation}
where
\begin{equation}
    F_{ij}(\omega) = x_i(\omega)x_j^*(\omega),\quad\text{for } i,j = 1,\,-1.
\end{equation}
For the diagonal terms, $F_{ii}(\omega)=|x_i(\omega)|^2$ are real-valued functions, which is the same as the linear acceleration case \eqref{transfer function}. By contrast, the cross transfer functions $F_{1,-1}(\omega)$ and $F_{-1,1}(\omega)$ have complex values. 

Notably, the cross-PSDs and their transfer functions satisfy the conjugate symmetry relations: $S_{a_1a_{-1}}(\omega)=S_{a_{-1}a_1}^*(\omega)$ and $F_{1,-1}(\omega)=F_{-1,1}^*(\omega)$. As a result, the contribution of cross correlation in \eqref{higher noise sigma_phi} is purely real and given by $2\text{Re}\left[S_{a_1a_{-1}}(\omega)F_{1,-1}(\omega)\right]$.

For a special case that $\delta a_1(t)$ and $\delta a_{-1}(t)$ arise from a noise term $\delta L=mf(t)x^n$ with a Gaussian noise $f(t)$ by defining $\delta a_{\pm1}(t)\equiv f(t)x_{\pm1}^{n-1}(t)$, then the dephasing of the single arm becomes 
\begin{equation}\label{higher dephasing}
    \delta\phi_{\rm diff} = \frac{m}{\hbar}\int_0^{t_f} f(t)x^n(t)\mathrm{d}t.
\end{equation}
Thus, the variance of the differential phase also has a form \eqref{dephasing factor and transfer function} with a transfer function
\begin{equation}\label{higher transfer function}
    F(\omega)=|x_1^n(\omega)-x_{-1}^n(\omega)|^2.
\end{equation}
For the case $n=2$, i.e. the term $m\delta\omega^2(t)\hat{x}^2$ can be used to study various noises such as the gravity gradient noise and rotation noise, of which the dephasing effect has been studied in previous researches~\cite{PhysRevResearch.3.023178, PhysRevD.111.064004}.

\subsubsection*{Position Localisation Decoherence and Humpty-Dumpty Problem}

The fidelity defined in \eqref{fidelity between two paths} characterises the position localisation decoherence and the Humpty-Dumpty problem, which can be computed as
\begin{equation}
    C = \text{Tr}[\hat{D}(\Delta\alpha(t_f)\mathrm{e}^{i\omega_0 t_f})\hat{\rho}_0]\mathrm{e}^{i\text{Im}(\alpha_1(t_f)\alpha^*_{-1}(t_f))},
\end{equation}
where $\Delta\alpha(t_f)\equiv\alpha_1(t_f)-\alpha_{-1}(t_f)$. Specially, if the initial state is prepared as a Gaussian state, then the fidelity is 
\begin{equation}\label{contrast loss for path-dependent noise}
    C = \text{Tr}[\hat{D}(\Delta\alpha(t_f)\mathrm{e}^{i\omega_0 t_f})\hat{\rho}_0] = \mathrm{e}^{-\frac{|\Delta\alpha(t_f)|^2}{2}} = \mathrm{e}^{-\frac{(\Delta x)^2}{m\omega_0/\hbar}-\frac{(\Delta p)^2}{m\hbar\omega_0}}.
\end{equation}
Since both $\delta\alpha_{\pm1}$ follow Gaussian distribution, then $\delta\alpha_1-\delta\alpha_{-1}$ also follow Gaussian distribution with a variance $\sigma^2_{\rm tot}=\sigma_1^2+\sigma_{-1}^2$ if they are assumed independent. Note that $\sigma_{\pm1}$ are exactly the same, then the total variance is simply $\sigma^2_{\rm tot}=2\sigma_\alpha^2$, where $\sigma_\alpha^2$ is exactly the variance of a single arm given by \eqref{sigma alpha t} which is a linear response to the acceleration noise $\delta a(t)$ with a transfer function $\text{sinc}^2(\omega-\omega_0)t/2$. Thus, the ensemble average of the contrast is
\begin{equation}
    -\mathbb{E}[\log C] = \frac{1}{2}\mathbb{E}\left[|\delta\alpha_1-\delta\alpha_{-1}|^2\right] = \sigma_\alpha^2.
\end{equation}

\subsubsection*{Witness Loss}

To sum up, higher-order noises can cause both dephasing and contrast loss on the interferometer, and both effects are linear responses to $\delta a(t)$. Both effects can contribute a decay factor on off-diagonal terms of the spin space, so the observable result for the Ramsey interferometry becomes
\begin{equation}
    \mathrm{Tr}\left[\mathbb{E}[\hat{\rho}_{\rm spin}(t_f)]\hat{W}\right] = \frac{1}{2}\left[1+\mathrm{e}^{-\frac{\sigma_{\phi_{\rm diff}}^2}{2}}\mathrm{e}^{-\sigma_{\alpha}^2}\cos(\phi_{\rm diff})\right],
\end{equation}
and the corresponding witness loss is
\begin{equation}
\begin{split}
    \Delta W &= \frac{\mathrm{e}^{-\sigma_{\alpha}^2-\frac{\sigma_{\phi_{\rm diff}}^2}{2}}-1}{2}\cos\phi_{\rm diff} \\
        &\approx -\left(\sigma_{\alpha}^2+\frac{\sigma_{\phi_{\rm diff}}^2}{2}\right)\cos\phi_{\rm diff}.
\end{split}
\end{equation}
Recall that both $\sigma_{\alpha}^2$ and $\sigma_{\phi_{\rm diff}}^2$ are determined by the PSD at $\omega_0$, given by \eqref{sigma phi} and \eqref{sigma alpha tf}. The resulting loss of witness is therefore expressed as
\begin{equation}
\begin{aligned}
    \Delta W \approx -\left(\frac{2\pi^2Nm}{\hbar\omega_0^2} + \frac{2\pi m^2N^2(\Delta A)^2}{\omega_0\hbar^2}\right)S_{aa}(\omega_0)\cos\phi_{\rm diff} \\
        = -\left(\frac{\pi m }{\hbar\omega_0}t_f + \frac{m^2(\Delta A)^2\omega_0}{2\pi\hbar^2}t_f^2\right)S_{aa}(\omega_0)\cos\phi_{\rm diff},
\end{aligned}
\end{equation}
where $\Delta A=2g_{\rm NV}\mu_B\eta/(m\omega_0^2)$ is the size of the interferometer. 

Here, the first term describes the combined influence of position localisation decoherence and the Humpty-Dumpty problem, which is characterised by the diffusion of the classical trajectories and scales linearly with the total time $\Delta W_1\propto t_f$. In contrast, the second term accounts for the dephasing effect, which exhibits a quadratic dependence on the total time, i.e. $\Delta W_2\propto t_f^2$. A further distinction between these contributions lies in their dependence on the interferometric path separation: only the dephasing relies on the magnitude of the ideal trajectories, characterised by $\Delta A$.

\section{Application to Typical Noises}\label{section 6}

As applications, I will briefly discuss the dephasing and contrast loss of two typical noises in SGIs: magnetic field noise and trap frequency noise. 

\subsection{Magnetic Field Noise}

One of the most significant noise is the fluctuation of the magnetic field. For simplicity, this paper focuses on the the magnetic gradient noise $\delta\eta(t)$, coupling to the test mass via the Hamiltonian
\begin{equation}
    \hat{H}_{\rm m}(t)=\mu_B\delta\eta(t)\hat{S}_z\hat{x}.
\end{equation}
Since this noise couples to both $\hat{S}_z$ and $\hat{x}$, it inherently affects both the spin space and the spatial degrees of freedom. Consequently, a complete analysis of the resulting decoherence requires solving the full spin-position dynamics, which poses a considerable mathematical challenge. 

At leading order, however, the spin states can be assumed to remain fixed along $s_z=\pm1$ states. In this approximation, the spins reduces to c-number, and the noise Hamiltonian takes a simplified form
\begin{equation}
    \hat{H}_{\rm m}=\pm\mu_B\delta\eta(t)\hat{x},
\end{equation}
which corresponds to an effective acceleration noise of the generic form given in \eqref{noisy Hamiltonian}. Thus, the decoherence arises sorely from the spatial degrees of freedom, and can be analyse using the previously established results on linear acceleration noise.

The phase fluctuation of each arm is given by \eqref{delta_phi}, then the differential phase is fluctuated by the magnetic noise as 
\begin{equation}
    \delta\phi_{\rm diff} = \frac{\mu_B}{\hbar}\int_0^{t_f}\delta\eta(t)(x_1(t)+x_{-1}(t))\mathrm{d}t.
\end{equation}
Hence, the variance of the phase fluctuation takes a similar form as the case for a uniform acceleration noise, given by \eqref{dephasing factor and transfer function}, where the transfer function takes a different form is the Fourier transform of the averaged position of two paths rather than the differential position between them. The variance can be formulated as
\begin{equation}
    \sigma_{\phi_{\rm diff}}^2 = \frac{m^2}{\hbar^2}\int S_{aa}(\omega)|x_1(t)+x_{-1}(t)|^2\mathrm{d}\omega.
\end{equation}
Specially, for a test mass drvien by a constant magnetic gradient without bias~\footnote{If there is a biased magnetic field $B_0\neq0$, then the amplitude $A_{\pm}$ defined in \eqref{trajectories for constant magnetic gradient} are not opposite with each other. As a result, the trajectories $x_1(t)$ and $x_{-1}(t)$ don't cancel with each other. In this case, there is still a dephasing effect, and the transfer function takes a similar form as \eqref{N transfer function} but the factor $\Delta A$ is replaced by $A_++A_-$.},
i.e. $B=\eta x$, the trajectories are given by \eqref{trajectories for constant magnetic gradient}. In this case, the trajectories $x_1(t)$ and $x_{-1}(t)$ are opposite with each other, that is,
\begin{equation}
    x_1(t) + x_{-1}(t) \equiv 0.
\end{equation}
Consequently, the phase fluctuation can cancel with each other, that is, $\delta\phi_{\rm diff}\equiv0$. In other words, the magnetic field noise does not cause dephasing effect via the spatial path integral phase.

Apart from inducing dephasing, the magnetic noise can also cause a non-closure problem, shown as Fig.\,\ref{Fig. uniform vs non-uniform}\,(c). Since the magnetic noise does not distort the shape of wavefunction, the contrast loss is entirely determined by the displacement in the classical phase space, given by \eqref{contrast loss for path-dependent noise}, i.e.
\begin{equation}
    C = \exp\left[-\frac{|\Delta\alpha(t_f)|^2}{2}\right],
\end{equation}
where $\Delta\alpha(t_f)\equiv\alpha_+(t_f)-\alpha_-(t_f)$ and the complex coordinates $\alpha_\pm(t_f)$ are defined by \eqref{alpha t}. Because the ideal trajectories are designed to overlap perfectly, the displacement $\Delta\alpha(t_f)$ arises sorely from the noise contribution, given by
\begin{equation}
    \Delta\alpha(t_f) = 2ig_{\rm NV}\mu_B\sqrt{\frac{m}{2\hbar\omega_0}}\mathrm{e}^{-i\omega_0t_f}\int_0^{t_f}\delta\eta(t)\mathrm{e}^{i\omega_0t}\,\mathrm{d}t. 
\end{equation}
For the stationary noise, the ensemble-averaged contrast can be evaluated using the Wiener-Khinchin theorem as
\begin{equation}
    -\mathbb{E}[\log C] = \frac{mg_{\rm NV}^2\mu_B^2}{\hbar\omega_0}S_{\eta\eta}(\omega_0),
\end{equation}
which is sorely determined by the PSD $S_{\eta\eta}$ of the magnetic noise at the intrinsic frequency $\omega_0$ of the test mass.

\subsection{Quadratic Noise}

Another common noise coupling mechanism involves a quadratic interaction with the test mass, represented by the Hamiltonian $\hat{H}_{\rm noise}=m\delta\omega^2(t)\hat{x}^2$. This form of coupling can arise from fluctuations of the trapping potential, gravity gradients or rotational motion~\cite{PhysRevD.111.064004}. 

The associated dephasing can be formulated using \eqref{higher dephasing}, which again yields a linear-response relation with the transfer function given by \eqref{higher transfer function} with $n=2$, as previously analyzed in Refs.~\cite{PhysRevResearch.3.023178, PhysRevD.111.064004}.

In contrast to the linear-coupling case, this type of noise can induce a Humpty-Dumpty problem via both fluctuation of the trajectories and the wavefunction shapes. As is derived in Appendix \ref{appendix d}, the fluctuation of each trajectory can be computed by perturbation method, yielding the leading-order result
\begin{equation}\label{delta alpha quadratic}
    \delta\alpha_c(t) = -\frac{i}{\omega_0}\mathrm{e}^{-i\omega_0t}\int_0^t\delta\omega^2(\tau)\text{Re}[\alpha_c^{(0)}(\tau)]\mathrm{e}^{i\omega_0\tau}\,\mathrm{d}\tau,
\end{equation}
where $\alpha_c^{(0)}(\tau)$ is the ideal phase-space trajectory defined in \eqref{alpha t}.

In addition, the noise distorts the shape of the Wigner function of the test mass via squeezed operators. As is derived in Appendix \ref{appendix d}, the squeezed operator at the leading-order is given by
\begin{equation}\label{approximated squeezed operator}
    \hat{S}(\rho) \approx \exp\left[\frac{i}{2\hbar}\delta\rho(t)(\hat{x}\hat{p}+\hat{p}\hat{x})\right]\exp\left[\frac{i}{2\hbar}m\delta\dot{\rho}(t)\hat{x}^2\right],
\end{equation}
where the first exponential corresponds to a squeezing operation in phase space, and the second represents a shearing transformation. The scaling factor is $\rho(t)=1+\delta\rho(t)$, with the perturbative correction $\delta\rho(t)$ and its time derivative given by
\begin{equation}
\begin{aligned}
    \delta\rho(t) &= -\frac{1}{2\omega_0}\int_0^t \delta\omega^2(\tau)\sin[2\omega_0(t-\tau)]\,\mathrm{d}\tau, \\
    \delta\dot{\rho}(t) &= -\int_0^t \delta\omega^2(\tau)\cos[2\omega_0(t-\tau)]\,\mathrm{d}\tau.
\end{aligned}
\end{equation}

However, due to the unitarity of the squeezed operator $\hat{S}(\rho)$, it doesn't affect the overlap of the final states~\footnote{Suppose the final states before squeezing are $\ket{\psi_+}$ and $\ket{\psi_-}$. After applying the squeezing operator, their overlap becomes
\begin{equation*}
    \bra{\psi_+}\hat{S}^\dagger(\rho)\hat{S}(\rho)\ket{\psi_-} = \braket{\psi_+|\psi_-},
\end{equation*}
which remains unchanged because $\hat{S}(\rho)$ is unitary.
}. Therefore, the contrast loss of the final state is sorely determined by the displacement difference $\Delta\alpha(t) = \delta\alpha_+(t)-\delta\alpha_-(t)$, analogous to the magnetic-noise case. The resulting contrast is
\begin{equation}
    C = \exp\left[-\frac{|\Delta\alpha(t_f)|^2}{2}\right].
\end{equation}
The displacement difference at the final time is
\begin{equation}
    \Delta\alpha(t_f) = -\frac{i}{\omega_0}\mathrm{e}^{-i\omega_0t}\int_0^t\delta\omega^2(\tau)\text{Re}[\Delta\alpha^{(0)}(\tau)]\mathrm{e}^{i\omega_0\tau}\,\mathrm{d}\tau,
\end{equation}
where $\Delta\alpha^{(0)}(\tau)\equiv\alpha^{(0)}_+(\tau)-\alpha^{(0)}_-(\tau)$ represents the ideal differential trajectory, and the complex coordinates $\alpha_\pm(t_f)$ are defined by \eqref{alpha t}, which can be computed as
\begin{equation}
    \text{Re}[\Delta\alpha^{(0)}(\tau)] = \sqrt{\frac{m}{2\hbar\omega_0}}\int_0^\tau\Delta a(\tau')\sin(\omega_0(\tau-\tau'))\,\mathrm{d}\tau'.
\end{equation}
This result indicates that the displacement $\Delta\alpha(t_f)$ is a linear response to the noise $\delta\omega^2$. Consequently, the logarithm of the contrast is also a linear response, determined by the PSD $S_{\omega^2\omega^2}$ of $\delta\omega^2$ given by
\begin{equation}\label{linear response for contrast loss}
    -\mathbb{E}[\log C] = \frac{1}{2}\int S_{\omega^2\omega^2}(\Omega)F(\Omega)\,\mathrm{d}\Omega,
\end{equation}
with a transfer function
\begin{equation}
    F(\Omega) = \bigg|\frac{1}{\omega_0}\int_0^{t_f}\text{Re}[\Delta\alpha^{(0)}(\tau)]\mathrm{e}^{i\Omega\tau}\,\mathrm{d}\tau\bigg|^2.
\end{equation}
Here, the frequency variable in the PSD domain is denoted by the capital letter $\Omega$ to distinguish it from the intrinsic frequency $\omega_0$ of the test mass and its fluctuation $\delta\omega$. Note that the quantity $\text{Re}[\Delta\alpha^{(0)}(\tau)]$ is exactly proportional to $x_1(t)-x_{-1}(t)$ in \eqref{transfer function}, so this transfer function is essentially the same as the transfer function of the dephasing induced by linear acceleration noise, up to some numerical factors.

For the constant driven force $\Delta a(t)\equiv\Delta A$ as given in \eqref{trajectories for constant magnetic gradient}, and for the final time $t_f=2N\pi/\omega_0$, the transfer function reduces to 
\begin{equation}
    F(\Omega) = \frac{2m(\Delta A)^2}{\hbar\omega_0}\cdot\frac{\sin^2(N\pi\Omega/\omega_0)}{\Omega^2(\Omega^2-\omega_0^2)^2},
\end{equation}
which is the same as the transfer function \eqref{N transfer function} of the dephasing induced by linear acceleration noise as expect. Hence, as is derived in Appendix \ref{appendix c}, the contrast loss can be also evaluated using the residue theorem, yielding
\begin{equation}
    -\mathbb{E}[\log C] = \frac{mN^2\pi^2(\Delta A)^2}{\hbar\omega_0^6}S_{\omega^2\omega^2}(\omega_0).
\end{equation}

Comparing to the constrast loss induced by the quadratic noise and magnetic noise, one finds that both are proportional to the noise PSD evaluated at the intrinsic frequency $\omega_0$, reflecting a resonance between the noise and the test mass. 

However, the contrast loss due to magnetic-field noise is independent of the ideal trajectories, whereas the quadratic noise couples to the ideal trajectories and scales with the number of periods $N$. This property arises from their coupling Hamiltonian: magnetic-field noise couples linearly to $\hat{x}$ and is therefore insensitive to the differential trajectory, whereas quadratic noise couples through $\hat{x}^2$, making the contrast loss proportional to $\Delta x$.

\section{Disucssion and Summary}\label{section 7}

\subsection{Role of randomness}

In the end, I draw a remark on the role of randomness. This distinction can be quantitatively captured by the \emph{purity} $\text{Tr}[\hat{\rho}^2]$ or the \emph{von Neumann entropy} $S=-\text{Tr}\left[\hat{\rho}\ln\hat{\rho}\right]$. The purities with and without the ensemble average are
\footnote{The trick to compute the purity and the von Neumann entropy is to diagonalize $\hat{\rho}_{\rm spin}$. In particular, suppose the non-zero eigenvalues of $\hat{\rho}_{\rm spin}$ are $\{\lambda_i\}$, then the purity and the von Neumann entropy of $\hat{\rho}_{\rm spin}$ are $\sum_i\lambda_i^2$ and $-\sum_i\lambda_i\ln\lambda_i$. For the density matrix $\hat{\rho}_{\rm spin}$ with a decay factor $\mathrm{e}^{-\Gamma_{\rm tot}}$ on the off-diagonal terms, its eigenvalues are $\lambda_{1, 2}=(1\pm\mathrm{e}^{-\Gamma_{\rm tot}})/2$ and $\lambda_3=0$. Specially, when the total decoherence factor $\Gamma_{\rm tot}$ is very small, then the entropy can be approximated as
\begin{equation*}
\begin{split}
    S &= -\left[\frac{1+\mathrm{e}^{-\Gamma_{\rm tot}}}{2}\ln\frac{1+\mathrm{e}^{-\Gamma_{\rm tot}}}{2} + \frac{1-\mathrm{e}^{-\Gamma_{\rm tot}}}{2}\ln\frac{1-\mathrm{e}^{-\Gamma_{\rm tot}}}{2}\right] \\
        &\approx \frac{1+\mathrm{e}^{-\Gamma_{\rm tot}}}{2}\cdot\frac{1-\mathrm{e}^{-\Gamma_{\rm tot}}}{2} + \frac{1-\mathrm{e}^{-\Gamma_{\rm tot}}}{2}\cdot\frac{1+\mathrm{e}^{-\Gamma_{\rm tot}}}{2} \\
        &= \frac{1}{2}(1-\mathrm{e}^{-2\Gamma_{\rm tot}}) \approx \Gamma_{\rm tot},
\end{split}
\end{equation*}
where the first approximated equation used an approximation $\ln((1\pm\mathrm{e}^{-\Gamma_{\rm tot}})/2)\approx(-1\pm\mathrm{e}^{\Gamma_{\rm tot}})/2$.
}
\begin{equation}
\begin{aligned}
    \text{Tr}[\hat{\rho}_{\rm spin}^2(t_f)] &= \frac{1}{2}(1+C^2) \approx 1-(-\log C), \\
    \text{Tr}\left[\mathbb{E}[\hat{\rho}_{\rm spin}(t_f)]^2\right] &= \frac{1}{2}(1+\mathbb{E}[C]^2\mathrm{e}^{-\sigma_{\phi_{\rm diff}}^2}) \approx 1 - \frac{\sigma_{\phi_{\rm diff}}^2}{2} - \sigma_\alpha^2,
\end{aligned}
\end{equation}
and the corresponding entropies are
\begin{equation}
\begin{aligned}
    S[\hat{\rho}_{\rm spin}] &\approx -\log C, \\
    S[\mathbb{E}[\hat{\rho}_{\rm spin}]] &\approx \frac{\sigma_{\phi_{\rm diff}}^2}{2} + \sigma_\alpha^2.
\end{aligned}
\end{equation}

The expressions for purity and entropy show that differential phase fluctuation under a determinstic fluctuation does not affect the purity or the entropy. However, a stochastic noise leads to an exponential decay factor due to the ensemble average, resulting in dephasing, characterised by purity loss and entropy growth $\sigma_{\phi_{\rm diff}}^2/2$.

Likewise, spatial fluctuation under external determinstic perturbation results in contrast loss, where the total information remains but the spin information becomes harder to extract. By contrast, a stochastic noise leads to genuine position-localisation decoherence after ensemble average, where quantum information is irreversibly lost.

From the point of view of information, ensemble averaging of the random noise transforms reversible contrast loss into irreversible information loss. The information of the quantum system gets lost even if the environment is classical as long as the environment is stochastic. Thus, the information loss indicates a deep connection between the classical randomness and the quantumness~\cite{2302.10778, 2309.03085}.

\subsection{Summary and Outlook}

This paper presents a comprehensive theoretical analysis on the on the decoherence of a Stern-Gerlach interferometer induced by classical stochastic acceleration noise. Based on the exact dynamics of the spatial motion, I analyse the various decoherence effects in an SGI.  

Section 3 presents a rigorous proof on the linear response relationship between the spatial dephasing and the external noise, where the associated transfer function is the Fourier transform of the unperturbed trajectories. This section further demonstrates that the transfer function behaves approximately as a delta-function at the intrinsic frequency $\omega_0$ of the test mass, which can be physically interpreted as the resonance between the test mass and the external noise. 

Section 4 investigates additional potential decoherence mechanisms, including the perturbations of the Wigner function, non-closure of the classical trajectories and velocities, and the position localisation decoherence. This section concludes that, for a linear acceleration noise, these effects can cancel with each other and thus do not affect the witness result. 

For higher-order noise, the leading-order decoherence effects can be treated as a path-dependent acceleration noise. From this perspective, the spatial dephasing remains a linear response to the noise, although the transfer function is modified according to the noise Hamiltonian. In this case, the non-closure problem, together with the position localisation decoherence, leads to a contrast loss which affects the spin witness, characterised by the PSD at $\omega_0$, which is also a resonance between the test mass and the noise.

Based on the general discussion, Section 6 applies the theoretical framework on two representative noise sources, magnetic field noise and quadratic noise. For magnetic field noise, the spatial dephasing cancels out, while the contrast loss under ensemble average is characterised by the PSD at $\omega_0$. For a quadratic noise, the spatial dephasing is still a linear response with a modified transfer function, and again the contrast loss under ensemble average is characterised by the PSD at $\omega_0$.

In summary, this paper develops a theoretical framework for linear effects of generic stochastic classical noise. The theoretical analysis methods presented here are readily adaptable to experiments based on matter-wave interferometry, and several of the results and the associated physical intuition can be directly applied in such setups. Future theoretical investigation may explore the coupling between the spatial and spin degrees of freedom, and study effects such as spin dephasing, relaxation, and Landau-Zenor transition (Majorana flip)~\cite{IVAKHNENKO20231, Band_2022} induced by spatial noise.

\appendix

\section{Analytical Solution for the Schrodinger Equation with Acceleration Noises}\label{appendix a}

In this appendix, I will show the math details to solve the Schrodinger equation with the time-dependent Hamiltonian \eqref{simplified Hamiltonian} by the \emph{Lewis-Riesenfeld invariant} and \emph{quantum Arnold transformation} method~\cite{lewis1969exact, 10.1063/1.525329, A_K_Dhara_1984, Guerrero_2015, Aldaya_2011}.

The basic idea~\cite{MIZRAHI1989465} is to find a so-called \emph{Lewis-Riesenfeld operator} $\hat{I}(t)$ such that all the eigenvalues $\{\lambda_n\}$ of $\hat{I}(t)$ are time-independent and all the instantaneous eigenstates $\{\ket{n(t)}\}$ of $\hat{I}(t)$ are solutions of Schrodinger equation. Then the time-evolution operator is given by $\hat{U}(t)=\sum_{n}\mathrm{e}^{i\alpha_n(t)}\ket{n(t)}\bra{n(0)}$, where $\alpha_n(t)=\int_0^t\bra{n(t')}\left(i\mathrm{d}/\mathrm{d}t'-\hat{H}(t')/\hbar\right)\ket{n(t')}\mathrm{d}t'$ is known as the \emph{Lewis-Riesenfeld phase}~\cite{lewis1969exact, MIZRAHI1989465}. From the point view of differential geometry, the instantaneous eigenstates $\{\ket{n(t)}\}$ form a moving frame of the Hilbert space as a manifold, and the Lewis-Riesenfeld phase is a kind of geometric phase arising from the connection 1-form of the manifold. 

Such time-dependent operator $\hat{I}(t)$ can be generally constructed by a reference frame transformation on the Hamiltonian from the lab reference to the co-moving frame~\cite{Guerrero_2015}, known as the \emph{quantum Arnold transformation}~\cite{Aldaya_2011}. In particular, for a driven oscillator with a Hamiltonian \eqref{simplified Hamiltonian}, consider a translation operator in the phase space of the oscillator
\begin{equation}
\begin{aligned}
    \hat{\mathcal{U}}(t) &= \exp\left[-\frac{i}{\hbar}(\hat{x}p(t)-\hat{p}x(t))\right] = \hat{D}(-\alpha(t)), \\
    \alpha(t) &= \sqrt{\frac{m\omega_0}{2\hbar}}(x(t)+ip(t)/m\omega_0).
\end{aligned}
\end{equation}
Under the unitary transformation, the wavefunction changes as $\ket{\psi}\to\ket{\psi'}=\hat{\mathcal{U}}(t)\ket{\psi}$, while the Hamiltonian changes as 
\begin{equation}\label{H transformation}
    \hat{H}(t) \to \hat{\mathcal{U}}\hat{H}\hat{\mathcal{U}}^\dagger - i\hbar\hat{\mathcal{U}}\frac{\mathrm{d}}{\mathrm{d}t}\hat{\mathcal{U}}^\dagger.
\end{equation}
The term $-i\hbar\hat{\mathcal{U}}(\mathrm{d}\hat{\mathcal{U}}^\dagger/\mathrm{d}t)$, often referred as the \emph{counter-diabatic} term $\hat{H}_{CD}$~\cite{KOLODRUBETZ20171}, mathematically arises from the connection 1-form to keep the Schrodinger equation covariant under the transformation $\hat{\mathcal{U}}(t)$. If the trajectory $x(t)$ and momentum $p(t)$ is chosen satisfying the canonical equation defined by the Hamiltonian \eqref{simplified Hamiltonian}
\begin{equation}\label{EOM}
    \left\{
        \begin{aligned}
            \dot{x}_c &= \frac{p_c}{m} \\
            \dot{p}_c &= -m\omega_0^2x_c + ma(t),
        \end{aligned}
    \right.
\end{equation}
which is the real-valued form for \eqref{dynamical alpha}, then the Hamiltonian in the co-moving reference is 
\begin{equation}
    \hat{\mathcal{U}}\hat{H}\hat{\mathcal{U}}^\dagger - i\hbar\hat{\mathcal{U}}\frac{\mathrm{d}}{\mathrm{d}t}\hat{\mathcal{U}}^\dagger = \frac{\hat{p}^2}{2m} + \frac{1}{2}m\omega_0^2\hat{x}^2 + \delta(t),
\end{equation}
where the phase $\delta(t)$ is given by
\begin{equation}\label{L-R phase}
    \delta(t) = -\frac{1}{2}ma(t)x_c(t). 
\end{equation}
Therefore, the Hamiltonian in the comoving reference frame is time-independent except for a pure phase term $\delta(t)$. So the time-evolution of the wave function under the comoving reference is simply
\begin{equation}
    \ket{\psi'(t)} = \mathrm{e}^{-\frac{i}{\hbar}\int_0^t\delta(t')\mathrm{d}t'}\mathrm{e}^{-\frac{i}{\hbar}\left(\frac{\hat{p}^2}{2m} + \frac{1}{2}m\omega_0^2\hat{x}^2\right)t}\ket{\psi'(0)}.
\end{equation}
Then the wavefunction in the original reference frame is 
\begin{equation}\label{solution}
    \ket{\psi(t)} = \mathrm{e}^{-\frac{i}{\hbar}\int_0^t\delta(t')\mathrm{d}t'}\hat{\mathcal{U}}^\dagger(t)\mathrm{e}^{-\frac{i}{\hbar}\left(\frac{\hat{p}^2}{2m} + \frac{1}{2}m\omega_0^2\hat{x}^2\right)t}\hat{\mathcal{U}}(0)\ket{\psi(0)}.
\end{equation}
Furthermore, the equation of motion \eqref{EOM} allows to rewrite the phase $\delta(t)$ in another form
\begin{equation}\label{Lagrangian phase}
\begin{split}
    \delta(t) =& -\left(\frac{1}{2}m\dot{x}_c^2(t) - \frac{1}{2}m\omega_0^2x_c^2(t) + ma(t)x_c(t)\right) \\
        &+ \frac{1}{2}m\frac{\mathrm{d}}{\mathrm{d}t}(\dot{x}_cx_c).
\end{split}
\end{equation}
This means that the $\delta(t)$ phase term is exactly the minus Lagrangian $L[x_c(t),\dot{x}_c(t)]$ of the system \eqref{simplified Hamiltonian} along the classical trajectory up to a total derivative term. Thus, the global phase in \eqref{solution} is exactly the path integral along the classical trajectory $\phi=\int_0^tL[x_c(t'),\dot{x}_c(t')]/\hbar\mathrm{d}t'$, so \eqref{solution} presents the time evolution operator \eqref{general solution}.
Remarkably, since the $\delta(t)$ phase arises from the gauge term $-i\hbar\hat{\mathcal{U}}(\mathrm{d}\hat{\mathcal{U}}^\dagger/\mathrm{d}t)$, the Feynman's path integral phase in this system has a physical interpretation of geometric phase.

A final remark is that a total derivative term in $\delta(t)$ is usually tolerant in an interferometry experiment because the initial and final position and momentum of different trajectories are the same and they will cancel with each other. It is notable that such a total derivative term is related to the choice of the unitary transform operator $\hat{\mathcal{U}}(t)$. In particular, the operator $\hat{\mathcal{U}}(t)$ allows an arbitrary phase and can be chosen as, for example~\cite{hogan2009light}, 
\begin{equation}
    \hat{\mathcal{U}}'(t) = \mathrm{e}^{\frac{i}{\hbar}\hat{p}x_c(t)}\mathrm{e}^{-\frac{i}{\hbar}\hat{x}p_c(t)} = \hat{\mathcal{U}}(t)\mathrm{e}^{\frac{i}{2\hbar}p_c(t)x_c(t)}.
\end{equation}
The corresponding pure phase term $\delta(t)$ for $\hat{\mathcal{U}}'(t)$ is~\cite{doi:10.1139/p58-038} 
\begin{equation}
    \delta'(t) = \frac{1}{2}m\dot{x}_c^2(t) - \frac{1}{2}m\omega_0^2x_c^2(t).
\end{equation}
Based on the equation of motion \eqref{EOM}, one can easily check the equivalence between $\delta(t)$ and $\delta'(t)$ up to a total derivative term $\frac{1}{2}m\frac{\mathrm{d}}{\mathrm{d}t}(x_c\dot{x}_c)$, i.e.
\begin{equation}
    -\frac{1}{2}ma(t)x_c(t) = \frac{1}{2}m\dot{x}_c^2(t) - \frac{1}{2}m\omega_0^2x_c^2(t) - \frac{1}{2}m\frac{\mathrm{d}}{\mathrm{d}t}(x_c\dot{x}_c).
\end{equation}

\section{Witness for Ramsey Interferometry in Spin-1 System}\label{appendix b}

In this appendix, the witness for Ramsey interferometry in spin-1 system will be constructed. A Ramsey interferometry is to measure the population of the spin state $\ket{S_z=0}$ after acting an external microwave pulse on it.
The Hamiltonian for the Rabi oscillation of the spin-1 system under the microwave pulse is given by $\hat{H}_p=\hbar\Omega_p\left(\ket{1}\bra{0}+\ket{-1}\bra{0}+h.c.\right)$, where $h.c.$ stands for "Hermitian conjugate", and $\Omega_p$ is the frequency of the microwave. The Hamiltonian can be written in a matrix form and can be diagonalized as
\begin{equation}
    \hat{H}_p = \hbar\Omega_p\begin{pmatrix}
        0 & 1 & 0 \\
        1 & 0 & 1 \\
        0 & 1 & 0
    \end{pmatrix} = \hbar\Omega_p\hat{T}^\dagger\begin{pmatrix}
        \sqrt{2} & 0 & 0 \\
        0 & 0 & 0 \\
        0 & 0 & -\sqrt{2}
    \end{pmatrix}\hat{T},
\end{equation}
where the transform matrix $\hat{T}$ is
\begin{equation}
    \hat{T} = \begin{pmatrix}
        \frac{1}{2} & \frac{1}{\sqrt{2}} & \frac{1}{2} \\
        \frac{1}{\sqrt{2}} & 0 & -\frac{1}{\sqrt{2}} \\
        \frac{1}{2} & -\frac{1}{\sqrt{2}} & \frac{1}{2}
    \end{pmatrix}.
\end{equation}
Then the time-evolution operator $\hat{U}_p(t)=\mathrm{e}^{-i\hat{H}_pt/\hbar}$ for the pulse acting on the spin-1 system is given by
\begin{equation}
\begin{split}
    \hat{U}_p(t) &= \hat{T}^\dagger \begin{pmatrix}
        \mathrm{e}^{-i\sqrt{2}\Omega_pt} & 0 & 0 \\
        0 & 1 & 0 \\
        0 & 0 & \mathrm{e}^{i\sqrt{2}\Omega_pt}
    \end{pmatrix} \hat{T} \\
    &= \begin{pmatrix}
        \cos^2\frac{\sqrt{2}\Omega_pt}{2} & -\frac{i}{\sqrt{2}}\sin\sqrt{2}\Omega_pt & -\sin^2\frac{\sqrt{2}\Omega_pt}{2} \\
        -\frac{i}{\sqrt{2}}\sin\sqrt{2}\Omega_pt & \cos\sqrt{2}\Omega_pt & -\frac{i}{\sqrt{2}}\sin\sqrt{2}\Omega_pt \\
        -\sin^2\frac{\sqrt{2}\Omega_pt}{2} & -\frac{i}{\sqrt{2}}\sin\sqrt{2}\Omega_pt & \cos^2\frac{\sqrt{2}\Omega_pt}{2}
    \end{pmatrix}.
\end{split}
\end{equation}
If the time of the microwave pulse is set as $t_p=\pi/(2\sqrt{2}\Omega_p)$, then the corresponding time-evolution operator is
\begin{equation}
    \hat{U}_{\pi/2} \equiv \hat{U}_p\left(t_p=\frac{\pi}{2\sqrt{2}\Omega_p}\right) = \begin{pmatrix}
        \frac{1}{2} & -\frac{i}{\sqrt{2}} & -\frac{1}{2} \\
        -\frac{i}{\sqrt{2}} & 0 & -\frac{i}{\sqrt{2}} \\
        -\frac{1}{2} & -\frac{i}{\sqrt{2}} & \frac{1}{2}
    \end{pmatrix}.
\end{equation}
After the $\pi/2$-pulse, the witness to measure the population of $\ket{S_z=0}$ is simply $\hat{W}'=\ket{S_z=0}\bra{S_z=0}$. Since the $\pi/2$-pulse makes a transformation on $\hat{\rho}_{\rm spin}$ to $\hat{U}_{\pi/2}\hat{\rho}_{\rm spin}\hat{U}_{\pi/2}^\dagger$, the observable result for the witness $\hat{W}'$ is $\mathrm{Tr}\left[\hat{W}'\hat{U}_{\pi/2}\hat{\rho}_{\rm spin}\hat{U}_{\pi/2}^\dagger\right]$. Therefore, the witness for Ramsey interferometry is given by
\begin{equation}
    \hat{W} = \hat{U}_{\pi/2}^\dagger\hat{W}'\hat{U}_{\pi/2} = \begin{pmatrix}
        \frac{1}{2} & 0 & \frac{1}{2} \\
        0 & 0 & 0 \\
        \frac{1}{2} & 0 & \frac{1}{2}
    \end{pmatrix}.
\end{equation}

\section{Computation of Integral via Residue Theorem}\label{appendix c}

In this paper, integrals of the form
\begin{equation}\label{integral I0}
    I_0 = A\int S(\omega)\frac{\sin^2(N\pi\omega/\omega_0)}{\omega^2(\omega^2-\omega_0^2)^2},
\end{equation}
or similar form appear repeatedly. Examples includes the dephasing \eqref{dephasing factor and transfer function} induced by linear acceleration noise in section 3, the phase-space diffusion \eqref{sigma2_alpha} of a single arm under noise in section 4, and the contrast loss \eqref{linear response for contrast loss} induced by quadratic noise in section 6. This appendix aims to evaluate this integral by the residue theorem. 

First of all, ignoring the amplitude factor $A$, this integral is exactly the real part of the integral
\begin{equation}\label{integral I}
    I = \int S(\omega)\frac{1- \mathrm{e}^{2\pi iN\omega/\omega_0}}{\omega^2(\omega^2-\omega_0^2)^2}\,\mathrm{d}\omega \equiv \int f(\omega)\,\mathrm{d}\omega,
\end{equation}
where the integrand is denoted as $f(\omega)$ for ease of writing. The poles of $f(\omega)$ consist of $0$, $\pm\omega_0$ and the poles of $S(\omega)$. 
Noticing that $S(\omega)$ is a real-valued even function on the real axis, the poles of $S(\omega)$ are symmetric to both the real axis and the imaginary axis, shown as Fig.\ref{poles of integrand}. The proof of this statement is summarised at the end of this appendix.
\begin{figure}
    \centering
    \begin{tikzpicture}[scale=0.8]
        \draw[->, thick, gray] (-4,0)--(4,0) node[anchor=west, black] {Re($\omega$)};
        \draw[->, thick, gray] (0,-2.5)--(0,4) node[anchor=south east, black] {Im($\omega$)};

        \fill (0,0) circle (0.05) node[anchor=north west] {$\omega=0$};
        \fill (1.5,0) circle (0.05) node[anchor=south] {$\omega_0$}; 
        \fill (-1.5,0) circle (0.05) node[anchor=south] {$-\omega_0$};
        \fill (3,0.5) circle (0.05) node[anchor=south] {$\omega_1$}; 
        \fill (3,-0.5) circle (0.05) node[anchor=north] {$\omega_1^*$};
        \fill (-3,0.5) circle (0.05) node[anchor=south] {$-\omega_1^*$}; 
        \fill (-3,-0.5) circle (0.05) node[anchor=north] {$-\omega_1$};
        \fill (1,1) circle (0.05) node[anchor=south] {$\omega_2$}; 
        \fill (1,-1) circle (0.05) node[anchor=north] {$\omega_2^*$};
        \fill (-1,1) circle (0.05) node[anchor=south] {$-\omega_2^*$}; 
        \fill (-1,-1) circle (0.05) node[anchor=north] {$-\omega_2$};
        \fill (2,1.5) circle (0.05) node[anchor=south] {$\omega_3$}; 
        \fill (2,-1.5) circle (0.05) node[anchor=north] {$\omega_3^*$};
        \fill (-2,1.5) circle (0.05) node[anchor=south] {$-\omega_3^*$}; 
        \fill (-2,-1.5) circle (0.05) node[anchor=north] {$-\omega_3$};
        
        \draw[green] (-3.75,0.05)--(3.75,0.05);
        \draw[green] (3.75,0) arc (0:180:3.75);
    \end{tikzpicture}
    \caption{\small Poles of the integrand in \eqref{integral I} and the integral path in the complex plane. The poles $\pm\omega_0$ arise from the transfer function $F(\omega)$, while other poles arise from the PSD $S(\omega)$.}
    \label{poles of integrand}
\end{figure}

According to the residue theorem, this integral equals the sum of the residues in the upper half of the complex plane, including those on the real axis. As is shown in Fig.\,\ref{poles of integrand}, $\omega_j$ and $-\omega^*_j$ are in the upper half of the complex plane, while $0$ and $\pm\omega_0$ are located on the real axis, so the integral \eqref{integral I} equals to
\begin{equation}
\begin{split}
    I &= 2\pi i\left[\sum\limits_{j}\left(\mathop{\text{Res}}\limits_{\omega=\omega_j}f(\omega) + \mathop{\text{Res}}\limits_{\omega=-\omega_j^*}f(\omega)\right)\right] \\
        &+ \pi i \left(\mathop{\text{Res}}\limits_{\omega=0}f(\omega) + \mathop{\text{Res}}\limits_{\omega=\omega_0}f(\omega) + \mathop{\text{Res}}\limits_{\omega=-\omega_0}f(\omega)\right).
\end{split}
\end{equation}

It is not difficult to check that $f(-\omega^*)=[f(\omega)]^*$~\footnote{
One can directly write
\begin{equation*}
    f(-\omega^*) = S(-\omega^*)\frac{1-\mathrm{e}^{-2\pi iN\omega^*/\omega_0}}{\omega^{*2}(\omega^{*2}-\omega_0^2)^2}.
\end{equation*}
It is obvious that $1-\mathrm{e}^{-2\pi iN\omega^*/\omega_0}=(1-\mathrm{e}^{2\pi iN\omega/\omega_0})^*$ and $\omega^{*2}(\omega^{*2}-\omega_0^2)^2=[\omega^2(\omega^2-\omega_0^2)^2]^*$. Besides, according to the properties $S(-\omega)=S(\omega)$ and $S(\omega^*)=[S(\omega)]^*$, one can obtain $S(-\omega^*)=[S(\omega)]^*$. Combine these equations, one may find $f(-\omega^*)=[f(\omega)]^*$. 
}, then the residue of $f(\omega)$ at the pole $-\omega_j^*$ satisfies 
\begin{equation}
    \mathop{\text{Res}}\limits_{\omega=-\omega_j^*}f(\omega) = \mathop{\text{Res}}\limits_{\omega=-\omega_j^*}\left[f(-\omega^*)\right]^* = \left[\mathop{\text{Res}}\limits_{\omega=\omega_j}f(\omega)\right]^*.
\end{equation}
Therefore, $2\pi i\left(\mathop{\text{Res}}\limits_{\omega=\omega_j}f(\omega)+\mathop{\text{Res}}\limits_{\omega=-\omega_j^*}f(\omega)\right)$ contributes a pure imaginary number. Consequently, the summation of all the poles $\omega_j$ and $-\omega_j^*$ has no contribution to the original integral $I_0\propto\text{Re}I$. In other words, $I_0$ can be completely determined by the poles $\omega=0$ and $\pm\omega_0$.

According to the symmetry of $f(\omega)$, the residue values of the poles $\pm\omega_0$ are equal. Since both of them are 2nd-order poles, their contribution to the integral \eqref{integral I} can be directly evaluated as
\begin{equation}
\begin{split}
    2\pi i\mathop{\text{Res}}\limits_{\omega=\omega_0}f(\omega) &= 2\pi i\frac{\mathrm{d}}{\mathrm{d}\omega}\left[S(\omega)\frac{1- \mathrm{e}^{2\pi iN\omega/\omega_0}}{\omega^2(\omega+\omega_0)^2}\right]\bigg|_{\omega=\omega_0} \\
        &= \frac{\pi^2N^2}{\omega_0^5}S(\omega_0). 
\end{split}
\end{equation}

As for the other 2nd-order pole $\omega=0$, it contributes to the integral \eqref{integral I} as
\begin{equation}
\begin{split}
    \pi i\mathop{\text{Res}}\limits_{\omega=0}f(\omega) &= \pi i\frac{\mathrm{d}}{\mathrm{d}\omega}\left[S(\omega)\frac{1- \mathrm{e}^{2\pi iN\omega/\omega_0}}{(\omega^2-\omega_0^2)^2}\right]\bigg|_{\omega=0} \\
        &= \frac{2\pi^2N^2}{\omega_0^5}S(\omega=0).
\end{split}
\end{equation}

In summary, the original integral \eqref{integral I0} is
\begin{equation}
    I_0 = \text{Re}I = \frac{\pi^2N^2}{\omega_0}\left[S(\omega_0)+2S(\omega=0)\right]. 
\end{equation}

However, the PSD at zero-frequency is usually vanished (like a Lorentzian PSD) or divergent (like a $1/f$ noise), both cases don't have observable effect on the witness of the test mass. Hence, the terms related to $S(\omega=0)$ are dropped off in the maintext.

\subsubsection*{Proof of lemma on symmetry of poles}
Now I prove the lemma that $-\omega_j$ and $\pm\omega_j^*$ are poles of $S(\omega)$ if $\omega_j$ is a pole. Without loss of generality, suppose $\omega_j$ is a $k$th-order pole in the first quadrant of the complex plane, then $S(\omega)$ can be expanded as a Laurent series at $\omega_j$ as
\begin{equation}
    S(\omega) = \frac{S_0}{(\omega-\omega_j)^k} + \cdots.
\end{equation}
Since $S(-\omega)=S(\omega)$, then one can obtain another Laurent series of $S(\omega)$ as
\begin{equation}
    S(\omega) = S(-\omega) = \frac{(-1)^kS_0}{(\omega+\omega_j)^k} + \cdots,
\end{equation}
which indicates that $-\omega_j$ is also a $k$th-order pole of $S(\omega)$. On the other hand, noticing that $S(\omega)$ takes real values on the real axis, i.e. $S(\omega)\in\mathbb{R},\,\forall\omega\in\mathbb{R}$, the Schwarz reflection principle states that $S(\omega^*)=[S(\omega)]^*$. This property implies another Laurent series of $S(\omega)$ as
\begin{equation}
    S(\omega) = [S(\omega^*)]^* = \frac{S_0^*}{(\omega-\omega_j^*)^k} + \cdots,
\end{equation}
which indicates that $\omega_j^*$ is also a $k$th-order pole of $S(\omega)$. Now that both $-\omega_j$ and $\omega_j^*$ are poles of $S(\omega)$, one can directly obtain that $-\omega_j^*$ is also a pole of $S(\omega)$.

\section{Master Equation of Position localisation Decoherence of Single Path}\label{appendix d}

In this appendix, I will derive the master equation for the position localisation decoherence of a single path of the interferometer. The procedure is similar to the derivation of the \emph{Redfield master equation} which can be found in textbooks such as \cite{10.1093/acprof:oso/9780199213900.003.03}.

Let's work under the interaction picture where the total Hamiltonian \eqref{simplified Hamiltonian} of a single arm can be divided into $\hat{H}_0$ and $\hat{H}_{\rm int}(t)=-ma(t)\hat{x}$, where $\hat{H}_0$ is of a simple oscillator and $\hat{H}_{\rm int}(t)$ contains all force terms. So the driven force and the gravity signal perform as the mean value of $ma(t)$, while the acceleration noise affects the auto-correlation of $ma(t)$. The dynamical equation for $\hat{\rho}^I(t)$ is
\begin{equation}\label{EOM for rho I}
    \frac{\mathrm{d}}{\mathrm{d}t}\hat{\rho}^I = -\frac{i}{\hbar}[\hat{H}^I_{\rm int}(t), \hat{\rho}^I(t)].
\end{equation}
This differential equation can be directly written in an integral form as
\begin{equation}\label{integral for rho I}
    \hat{\rho}^I(t) = \hat{\rho}^I(0) - \frac{i}{\hbar}\int_0^t[\hat{H}^I_{\rm int}(t'), \hat{\rho}^I(t')]\mathrm{d}t'.
\end{equation}
The basic idea is to do recursion for $\hat{\rho}^I(t)$, that is, to insert \eqref{integral for rho I} to the right side of \eqref{EOM for rho I}, then one may obtain
\begin{equation}
    \frac{\mathrm{d}}{\mathrm{d}t}\hat{\rho}^I = -\frac{1}{\hbar^2}\int_0^t[\hat{H}^I_{\rm int}(t), [\hat{H}^I_{\rm int}(t'), \hat{\rho}^I(t')]]\mathrm{d}t',
\end{equation}
where $\hat{\rho}^I(0)$ is assumed to commute with $\hat{H}^I_{\rm int}(t)$, i.e. $[\hat{H}^I_{\rm int}(t), \hat{\rho}^I(0)]=0$. Besides, the decoherence is usually assumed memoryless, known as the \emph{Markovian} assumption~\cite{schlosshauer2007quantum, 10.1093/acprof:oso/9780199213900.003.03} which means that the $\hat{\rho}^I(t)$ can be assumed to only depend on $t$, i.e. the $\hat{\rho}^I(t')$ in the integral can be substituted by $\hat{\rho}^I(t)$.
In addition, one can substitute $t'$ by $t-\tau$ in the integral, and take the upper limit of the integral to infinite, then obtain 
\begin{equation}
    \frac{\mathrm{d}}{\mathrm{d}t}\hat{\rho}^I = -\frac{1}{\hbar^2}\int_0^\infty[\hat{H}^I_{\rm int}(t), [\hat{H}^I_{\rm int}(t-\tau), \hat{\rho}^I(t)]]\mathrm{d}\tau.
\end{equation}  
Further, by substituting the interaction Hamiltonian $\hat{H}^I_{\rm int}(t)=-ma(t)\hat{x}^I(t)$, and taking the ensemble average, the dynamical equation of the density matrix becomes to
\begin{equation}\label{Redfield master equation}
\begin{split}
    \frac{\mathrm{d}}{\mathrm{d}t}\mathbb{E}[\hat{\rho}^I(t)] =& -\frac{m^2}{\hbar^2}\int_0^\infty\mathbb{E}\left[a(t)a(t-\tau)\right] \\ 
        &\times \left[\hat{x}^I(t), \left[\hat{x}^I(t-\tau), \mathbb{E}[\hat{\rho}^I(t)]\right]\right]\mathrm{d}\tau.
\end{split}
\end{equation}
This is a deformation of the Redfield master equation for a classical stochastic noise. In the original Redfield equation, the environment is assumed as a quantum bath, so the density matrix of the test mass is obtained by tracing out the degrees of freedom of the quantum bath. For the classical random noise, the ensemble average plays a similar role as tracing out, which intimates a correspondence between randomness and quantumness~\cite{2302.10778, 2309.03085}. 

Note that the time-evolution of the position operator $\hat{x}^I(t)$ of a harmonic oscillator satisfies $\hat{x}^I(t-\tau)=\hat{x}^I(t)\cos\omega_0\tau-\hat{p}^I(t)/(m\omega_0)\sin\omega_0\tau\sim\hat{x}^I(t)\mathrm{e}^{i\omega_0\tau}$, so \eqref{Redfield master equation} becomes to
\begin{equation}
\begin{split}
    \frac{\mathrm{d}}{\mathrm{d}t}\mathbb{E}[\hat{\rho}^I(t)] =& -\frac{m^2}{\hbar^2}[\hat{x}^I(t), [\hat{x}^I(t), \mathbb{E}[\hat{\rho}^I(t)]]] \\
        &\times \int_0^\infty\mathbb{E}[a(t)a(t-\tau)]\mathrm{e}^{i\omega_0\tau}\mathrm{d}\tau.
\end{split}
\end{equation} 
Note that the integral is nothing but $2\pi S_{aa}(\omega_0)/2$ according to the Wiener-Khinchin theorem and the symmetry of $S_{aa}(\omega)$, i.e. $S_{aa}(-\omega)=S_{aa}(\omega)$. So this master equation describes a position-localisation decoherence with a dissipator $\Lambda=\pi m^2S_{aa}(\omega_0)/\hbar^2$, and it becomes to the master equation \eqref{master equation} in the main text by switching back to the Schrodinger picture.

\section{Perturbation Analysis on Quadratic Noise}\label{appendix e}

This appendix presents a detailed perturbative analysis on quadratic noise term $\hat{H}_{\rm noise}=m\delta\omega^2(t)\hat{x}^2$, as discussed in section \ref{section 5}.

The total Hamiltonian can be written as
\begin{equation}
    \hat{H}(t) = \frac{\hat{p}^2}{2m} + \frac{1}{2}m\omega^2(t)\hat{x}^2 - ma(t)\hat{x},
\end{equation}
where $\omega^2(t)=\omega_0^2+\delta\omega^2(t)$. The corresponding Lewis-Riesenfeld invariant is expressed as~\cite{10.1063/1.525329, A_K_Dhara_1984, PhysRevA.83.013415}
\begin{equation}
    \hat{I}(t) = \frac{\left(\rho(t)(\hat{p}-m\dot{x}_c)-m\dot{\rho}(\hat{x}-x_c(t))\right)^2}{2m} + \frac{1}{2}m\omega_0^2\left(\frac{\hat{x}-x_c(t)}{\rho(t)}\right)^2,
\end{equation}
where the auxiliary functions $\rho(t)$ and $x_c(t)$ satisfy the dynamical eqautions
\begin{equation}\label{Ermakov equation}
\begin{aligned}
    \ddot{x}_c + \omega^2(t)x_c(t) &= a(t), \\
    \ddot{\rho} + \omega^2(t)\rho(t) &= \frac{\omega_0^2}{\rho^3(t)}.
\end{aligned}
\end{equation}
Based on the L-R invariant, the time-evolution operator takes the form
\begin{equation}
    \hat{U}(t) = \mathrm{e}^{i\phi(t)}\hat{S}(\rho(t))\hat{D}(\alpha_c(t))\mathrm{e}^{-\frac{i}{\hbar}\hat{H}_0t}\hat{D}^\dagger(\alpha_c(0))\hat{S}^\dagger(\rho(0)),    
\end{equation}
where $\alpha_c(t)=\sqrt{m\omega_0/2\hbar}(x(t)+i\dot{x}(t)/\omega_0)$ is the complex coordinate of the classical phase space, and $\hat{S}$ is the squeeze operator given by 
\begin{equation}\label{squeezed operator}
    \hat{S}(\rho) = \exp\left[\frac{i}{2\hbar}\ln\rho(\hat{x}\hat{p}+\hat{p}\hat{x})\right]\exp\left[\frac{i}{2\hbar}m\rho\dot{\rho}\hat{x}^2\right].
\end{equation}
Compared to the solution of acceleration noise \eqref{general solution}, this time-evolution operator has an additional squeeze operator $\hat{S}$.

\subsubsection*{Perturbative solution of $x_c(t)$ and $\alpha_c(t)$}

The dynamical equation for $x_c(t)$ is given by the first equation of \eqref{Ermakov equation}. The perturbative solving method is similar as solving the Pinney equation \eqref{Pinney equation}. In particular, the solution can be written as a perturbative form 
\begin{equation}
    x_c(t)=x_c^{(0)}(t)+x_c^{(1)}(t),
\end{equation}
where the zero-th order $x_c^{(0)}(t)$ is exactly the real part of $\alpha_c(t)$ of an ideal interferometer, given by \eqref{alpha t}, i.e.
\begin{equation}
    x_c^{(0)}(t) = \frac{1}{\omega_0}\int_0^t\sin[\omega_0(t-\tau)]a(\tau)\,\mathrm{d}\tau.
\end{equation}
Then the first order perturbation $x_c^{(1)}(t)$ satisfy
\begin{equation}
    \ddot{x}_c^{(1)} + \omega_0^2x_c^{(1)}(t) = -\delta\omega^2(t)x_c^{(0)}(t).
\end{equation}
Then the first order perturbation solution is
\begin{equation}
    x_c^{(1)}(t) = -\frac{1}{\omega_0}\int_0^t\sin[\omega_0(t-\tau)]\delta\omega^2(\tau)x^{(0)}(\tau)\,\mathrm{d}\tau.
\end{equation}
Then one may compute the fluctuation of the complex coordinate $\delta\alpha_c(t)$ of classical phase space at the leading order as
\begin{equation}
    \delta\alpha_c(t) = -i\sqrt{\frac{m}{2\hbar\omega_0}}\mathrm{e}^{-i\omega_0t}\int_0^t\delta\omega^2(\tau)x_c^{(0)}(\tau)\mathrm{e}^{i\omega_0\tau}\,\mathrm{d}\tau,
\end{equation}
which is the same as \eqref{delta alpha}.

\subsubsection*{Perturbative solution of $\rho(t)$}

The second equation in the auxiliary dynamical equation \eqref{Ermakov equation} is known as the \emph{Ermakov equation}~\cite{Ermakov1880}. Its analytic solution takes the general form
\begin{equation}\label{Ermakov solution}
    \rho(t) = \sqrt{Ay_1^2(t)+2By_1(t)y_2(t)+Cy_2^2(t)},
\end{equation}
where $y_1(t)$ and $y_2(t)$ are two linear independent solutions of the so-called \emph{Pinney equation}~\cite{Pinney1950}
\begin{equation}\label{Pinney equation}
    \ddot{y} + \omega^2(t)y(t) = 0.
\end{equation}
The constants $A$, $B$ and $C$ in \eqref{Ermakov solution} are arbitrary but must satisty the constraint
\begin{equation}
    AC - B^2 = \frac{\omega_0^2}{(y_1\dot{y}_2-y_2\dot{y}_1)^2}.
\end{equation}
Since the Pinney equation \eqref{Pinney equation} does not admit a closed-form analytic solution in general, I will solve it by perturbation method. Suppose the solution of the Pinney equation \eqref{Pinney equation} can be expended as
\begin{equation}
    y(t)=y^{(0)}(t)+y^{(1)}(t),
\end{equation}
where the zero-th order $y^{(0)}(t)$ satisfies the static oscillator equation
\begin{equation}
    \ddot{y}^{(0)} + \omega_0^2(t)y^{(0)}(t) = 0,
\end{equation}
of which the two independent solutions are
\begin{equation}
    y_1^{(0)}(t) = \cos\omega_0t,\quad y_2^{(0)}(t) = \sin\omega_0t.
\end{equation} 
Substituting this expansion into the Pinney equation, the first-order correction $y^{(1)}(t)$ satisfies
\begin{equation}
    \ddot{y}^{(1)} + \omega_0^2y^{(1)}(t) = -\delta\omega^2(t)y^{(0)}(t).
\end{equation}
%
This inhomogeneous equation can be solved by Green's-function method. The solution is simply the convolution between the inhomogeneous term $-\delta\omega^2(t)y^{(0)}(t)$ and the Green's function $\sin\omega_0t$, given as
\begin{equation}
    y^{(1)}(t) = -\frac{1}{\omega_0}\int_0^t \delta\omega^2(\tau)\sin\omega_0(t-\tau)y^{(0)}(\tau)\,\mathrm{d}\tau.
\end{equation}
Hence, the perturbative solutions of the Pinney equation \eqref{Pinney equation} are
\begin{equation}\label{perturbated solution of Pinney}
\begin{aligned}
    y_1(t) = \cos\omega_0t - \frac{1}{\omega_0}\int_0^t \delta\omega^2(\tau)\sin\omega_0(t-\tau)\cos\omega_0\tau\,\mathrm{d}\tau, \\
    y_2(t) = \sin\omega_0t - \frac{1}{\omega_0}\int_0^t \delta\omega^2(\tau)\sin\omega_0(t-\tau)\sin\omega_0\tau\,\mathrm{d}\tau.
\end{aligned}
\end{equation}
Using these two linearly independent perturbative solutions, the Wronskian at first order becomes
\begin{equation}
    y_1\dot{y}_2-y_2\dot{y}_1 = \omega_0.
\end{equation}
Consequently, the constraint on the constants in \eqref{Ermakov solution} reduces to
\begin{equation}
    AC - B^2 = 1.
\end{equation}
Moreover, in the unperturbed case $\delta\omega^2(t)=0$, the physically relevant solution of the Ermakov equation is the constant function $\rho(t)\equiv1$. This fixes the auxiliary constants to 
\begin{equation}
    A = C = 1, \quad B = 0.
\end{equation}
Thereofre, the solution of the Ermakov equation \eqref{Ermakov solution} can be simplified as
\begin{equation}
    \rho(t) = \sqrt{y_1^2(t)+y_2^2(t)}.
\end{equation}
Substituting the perturbative solutions \eqref{perturbated solution of Pinney} into this equation, retaining only first-order terms (discarding those quadratic in the integral of $\delta\omega^2(\tau)$), and simplyfing the trigonometric functions, the perturbated solution is finally obtained as
\begin{equation}\label{approximated rho}
\begin{split}
    \rho(t) &\approx \sqrt{1-\frac{1}{\omega_0}\int_0^t \delta\omega^2(\tau)\sin2\omega_0(t-\tau)\,\mathrm{d}\tau}, \\ 
        &\approx 1 - \frac{1}{2\omega_0}\int_0^t \delta\omega^2(\tau)\sin[2\omega_0(t-\tau)]\,\mathrm{d}\tau.
\end{split}
\end{equation}

It indicates that the fluctuation of $\rho(t)$ is exactly a convolution between the input noise and the kernal $\sin2\omega_0t$, i.e. $\delta\rho(t)=[\delta\omega^2(\tau)\ast\sin2\omega_0\tau](t)$, which indicates the linear response of $\rho(t)$ to the noise $\delta\omega^2(t)$.

Substituting the perturbative result \eqref{approximated rho} of $\rho(t)$ to the squeezed operator \eqref{squeezed operator}, the perturbative squeezed operator is given by \eqref{approximated squeezed operator} in the maintext.

\section*{Acknowledgements}
    I would like to thank the China Scholarship Council (CSC) for the financial support. I also thank Guan-Zhuo Yang for the beneficial discussions.

\end{document}